# Exploring the Role of Molecular Dynamics Simulations in Most Recent Cancer Research: Insights into Treatment Strategies

___________________________________________________________________

Reza Bozorgpour[1*]

___________________________________________________________________

[1] Department of Biomedical Engineering, University of Wisconsin-Milwaukee, Milwaukee, US

## Abstract

Cancer is a complex disease that is characterized by uncontrolled growth and division of cells. It involves a complex interplay between genetic and environmental factors that lead to the initiation and progression of tumors. Recent advances in molecular dynamics simulations have revolutionized our understanding of the molecular mechanisms underlying cancer initiation and progression. Molecular dynamics simulations enable researchers to study the behavior of biomolecules at an atomic level, providing insights into the dynamics and interactions of proteins, nucleic acids, and other molecules involved in cancer development. In this review paper, we provide an overview of the latest advances in molecular dynamics simulations of cancer cells. We will discuss the principles of molecular dynamics simulations and their applications in cancer research. We also explore the role of molecular dynamics simulations in understanding the interactions between cancer cells and their microenvironment, including signaling pathways, protein interactions, and other molecular processes involved in tumor initiation and progression. In addition, we highlight the current challenges and opportunities in this field and discuss the potential for developing more accurate and personalized simulations. Overall, this review paper aims to provide a comprehensive overview of the current state of molecular dynamics simulations in cancer research, with a focus on the molecular mechanisms underlying cancer initiation and progression.

## 1. Introduction

Cancer, a non-infectious ailment, stands as a significant contributor to human fatalities in the 21st century. According to findings from the World Health Organization (WHO), cancer emerges as the primary cause of premature death in individuals under the age of 70 in 91 out of the 172 countries examined [1]. In most cases, the atypical proliferation of cells within different organs of the body can lead to the development of cancer. Consequently, the classification of cancers is determined by the anomalous cell growth occurring in particular organ systems [2]. There exist over 100 distinct cancer types, with certain organs, such as the breast, skin, colon, and lungs, being particularly susceptible to cancer development [3-6], often resulting in fatal outcomes. The primary modes of treating cancer typically involve chemotherapy, radiotherapy, and surgical interventions [7].

[*]Correspondence should be addressed to Reza Bozorgpour, Department of Biomedical Engineering, University of Wisconsin-Milwaukee, Milwaukee, US. Email: bozorgp2@uwm.ed

Chemotherapy constitutes the therapeutic approach to combat cancer by employing chemical substances known for their toxicity to cancerous cells. While chemotherapeutic agents typically impede the proliferation of cancer cells, they inadvertently affect healthy cells as well, giving rise to undesirable side effects. Furthermore, the persistence of micro metastases of cancer stem cells and the occurrence of cellular senescence frequently contribute to cancer relapse and the ineffectiveness of chemotherapy [8, 9]. Radiotherapy employs radiation to curtail the proliferation of cancer cells. Nevertheless, this treatment method comes with its own set of challenges, as the radiation can cause harm to genetic materials and disrupt protein and enzyme pathways in healthy cells [10]. Surgery serves as a treatment approach reserved for the removal of cancer-affected tissues and organs in the advanced stages when other treatments have proven ineffective. A significant hurdle associated with surgical treatment is the potential reappearance of cancer following the removal of affected organs and tissues due to relapse, posing a substantial challenge [11]. Importantly, the non-targeted nature of conventional cancer treatment methods contributes significantly to these challenges. Therefore, there has been considerable research interest in developing therapeutic strategies that specifically target cancer cells at malignant sites.

Numerous innovative treatment strategies, including hyperthermia, cryosurgery, and combination therapies, have been developed to selectively target cancer cells. Hyperthermia, for instance, involves the application of heat to treat cancer cells, capitalizing on their susceptibility to elevated temperatures compared to normal cells [12]. While this approach has shown promise, it does induce heat stress in normal cells, resulting in low therapeutic margins that necessitate further research to enhance effectiveness.

In a different vein, magnetic hyperthermia has emerged as an alternative to traditional hyperthermia. It involves the introduction of magnetic drug particles into the body to specifically bind with cancer cells. These particles, when subjected to an external magnetic field, generate localized heating, thereby inhibiting the growth of cancer cells. This method boasts the advantage of precision in targeting cancer cells; however, it's important to note that the use of magnetic particles as drug molecules can potentially harm normal cells [13]. Innovations in cancer treatment have introduced cryosurgery as a promising approach, utilizing cryodrugs injected into the body to specifically target cancer cells. These cryo-drugs work by binding to cancer cells, after which the targeted cells are subjected to freezing temperatures in their natural location to impede their growth [14]. Nevertheless, it's important to note that freezing cancer cells can also impact surrounding healthy tissues and organs. To tackle this challenge, novel methods like nanocryosurgery have been under exploration, aiming to minimize collateral damage [15].

In the quest for more effective cancer treatments, researchers have been exploring combinational strategies that integrate modern approaches with traditional chemical methods [16]. One such avenue involves the utilization of phytochemicals as substitutes for chemically derived anti-cancer drugs. This substitution helps mitigate side effects in combinational cancer therapy, offering a more holistic and gentler approach to treatment [17].

A multitude of proteins have emerged as key players in the intricate landscape of cancer. These proteins range from the well-known tumor suppressor p53 to the notorious metastasis enabler S100A4. In contrast to the earlier days when drug development primarily targeted tumors as a whole, contemporary drug design has taken a sharp turn towards pinpointing specific proteins. These novel drugs exhibit a distinct advantage: they tend to induce fewer and less severe side effects. This advantage stems from their precision in targeting only those proteins intricately linked to the progression of cancer. Furthermore, the advent of affordable and widely accessible sequencing technologies, alongside a wealth of genetic markers, has paved the way for the emergence of personalized cancer treatments [18]. This personalized approach tailors therapeutic interventions to the unique genetic makeup of each patient, offering promising prospects for more effective cancer management. However, it's essential to acknowledge that our understanding of the biology underlying many cancer-related proteins and mutations remains incomplete. Frequently, a marker associated with a specific type of cancer is identified, but the precise connection between that marker and the disease remains elusive. In essence, while we are making significant strides in the realm of cancer research and treatment, there is still much to uncover about the intricacies of these proteins and their roles in the onset and progression of cancer.

A profound understanding of a specific protein's functions and the impact of various mutations on its behavior often comes to light once the protein's structure is deciphered. However, it's crucial to recognize that proteins are far from static entities; it is their dynamic behavior that ultimately determines their roles in biological processes [19]. To unravel these dynamic nuances, various biophysical techniques such as Nuclear Magnetic Resonance (NMR), Förster Resonance Energy Transfer (FRET), X-ray Laue diffraction, Extended X-ray Absorption Fine Structure (EXAFS), and others have been employed.

These techniques, though powerful, are not without their limitations. They may face challenges concerning sensitivity, applicability to certain types of molecules, and the time scales over which they can capture dynamic changes. This is where computer-aided studies step in to complement experimental approaches. Molecular modeling, in particular, has emerged as a valuable tool in understanding protein dynamics, and its relevance extends into various clinically-focused research areas, such as amyloid-related diseases [20]. These computational methods enable researchers to explore intricate details that may not be readily accessible through experimental means. However, it's worth noting that the accessibility of these computational methods to cancer researchers can sometimes be hindered by a lack of understanding regarding their potential applications, advantages, and limitations. Bridging this knowledge gap is essential to harness the full potential of molecular modeling and computational approaches in cancer-related research, where insights into protein dynamics can be of paramount importance.

This review article delves into the realm of molecular modeling within the context of cancer research. Its primary objective is to provide a clear and accessible introduction to molecular modeling, presenting some of the commonly employed techniques in a manner understandable to those who may not be experts in the field. It's important to note that molecular modeling is just one facet of the broader field of computational biology, and this article's focus is specifically on this aspect. While the world of cancer research is vast and encompasses various computational approaches, including bioinformatics and mathematical modeling [21–24], this review article concentrates solely on molecular modeling methodologies. These methods involve the creation of computational models that simulate the behavior of molecules at a molecular or atomic level, offering valuable insights into the intricate processes underlying cancer development and progression. By narrowing its focus in this way, the article aims to provide a comprehensive yet accessible overview of a specific subset of tools and techniques used in cancer research, shedding light on how molecular modeling can contribute to our understanding of this complex disease.

## 2. Methods of molecular modelling and simulation

"Molecular modeling" has evolved significantly from its early beginnings. Many of us might recall those plastic balls and sticks that were used in chemistry classes to illustrate how atoms come together to form molecules. These physical models did find applications in research [25] but have largely given way to computer-generated models. Today, when we mention "molecular modeling," we're referring to the utilization of computer-generated models in the study of molecules, ranging from small atomic structures to complex biomolecules. These computer-generated models play a pivotal role in simulating processes that span an incredible range of time scales, from incredibly rapid events lasting mere femtoseconds ($10^{-15}$ seconds) to much slower processes that may take several seconds. It's crucial to recognize that the accuracy and level of detail provided by these models depend on the size and time frame of the system being studied. For instance, molecular modeling can discern sub-Ångström differences (less than a billionth of a meter) between structures, which can have a profound impact on how a drug molecule binds to its target receptor. On the other end of the spectrum, when dealing with large protein complexes that are orders of magnitude larger, a different set of modeling methods is required. In essence, the choice of modeling method is highly contingent on the specific problem under investigation. In the world of molecular modeling, a fundamental trade-off exists: the more accurate the method used, the greater the time and computational resources required to generate meaningful results, especially when comparing systems of similar size. Consequently, researchers must carefully select the modeling approach that aligns with the unique characteristics of their research problem. This adaptability allows molecular modeling to be a versatile and indispensable tool in scientific investigations across a wide spectrum of scientific disciplines.

In the realm of molecular modeling, while it's feasible to model collections of molecules, such as multiple peptides and lipids [26], the majority of studies in the domain of cancer research tend to focus on atomistic details. These methods, however, come with certain limitations. They are primarily applicable to molecules for which structural data have been experimentally determined through techniques like X-ray crystallography or Nuclear Magnetic Resonance (NMR), or molecules that can be accurately modeled using computational approaches (as elaborated below). Atomistic modeling is particularly limited in terms of both the size and timescale of the systems it can effectively address. For instance, contemporary techniques allow the study of soluble proteins on the order of $10^{-7}$ to $10^{-6}$ seconds, enabling insights into specific domain movements within a protein. However, they often fall short when it comes to examining more extensive conformational changes, such as protein folding or the activation of ion channels. While it's still possible to model longer processes and larger molecular complexes, it necessitates the application of advanced and somewhat approximate computational methods.

This article proceeds to discuss various molecular modeling approaches and their applications, breaking them down into four categories for clarity: atomistic simulation and modeling methods, modeling of proteins and protein complexes, modeling of interactions between proteins and drugs, and simplified approaches that do not provide atomistic-level details. These categories are somewhat artificial, with some methods overlapping or belonging to multiple categories. In such cases, the categorization reflects where each method is most commonly employed or where it holds the potential for significant utility in cancer research. Each method is introduced with an executive summary outlining its strengths, limitations, and general applicability, followed by a concise yet informative description. For visual clarity, a graphical representation of some of the most frequently used methods is provided in Figure 1.

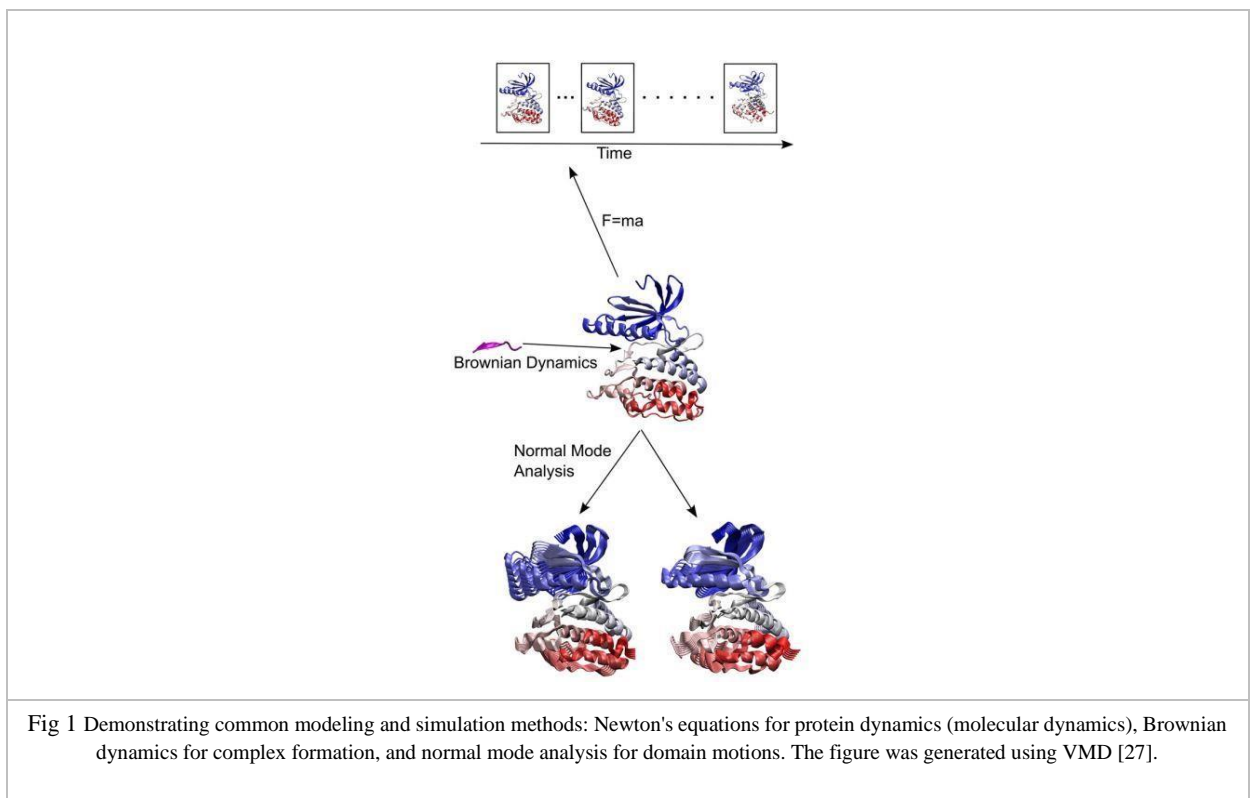

Fig 1 Demonstrating common modeling and simulation methods: Newton's equations for protein dynamics (molecular dynamics), Brownian dynamics for complex formation, and normal mode analysis for domain motions. The figure was generated using VMD [27].

In the subsequent sections of the article, several cancer-related studies employing these modeling methods will be explored, shedding further light on the successful applications of molecular modeling in the field of cancer research.

## 2.1. Atomistic simulation and modelling methods

Molecular simulation methods are essential for studying dynamic processes that are often challenging to observe experimentally, particularly in the context of biomacromolecules. These methods can also be applied to larger systems but require atomic-resolution structural information. Ideally, this structural data should originate from high-resolution techniques like protein crystallography or NMR. However, in cases where experimental data is unavailable or incomplete, computational models can serve as a substitute source of structural information. Molecular simulations enable a deep understanding of molecular dynamics, providing insights into complex biological systems at the atomic level.

### 2.1.1. Atomistic molecular dynamics simulations

#### 2.1.1.1. Advantages

This method is relatively user-friendly and provides dynamic data on atomic and molecular movements at an atomistic level, akin to watching a protein in an animated movie. It offers the convenience of extracting various structural observables from the simulations. Additionally, protein modeling using this approach often achieves a high level of realism, enhancing its utility in research and analysis.

#### 2.1.1.2. Constrains

When working with this method, its effectiveness is notably diminished in cases where there is a paucity of experimental data concerning the biomolecule under scrutiny. Furthermore, it is not well-suited for the examination of extensive or prolonged processes without incorporating supplementary modifications, which may result in decreased precision. Additionally, this method is not inherently suitable for investigating processes that entail the breaking or formation of covalent bonds, necessitating specialized adjustments for such scenarios.

### 2.1.1.3. Practical use

Let's say you have a protein's structure bound to a drug and want to explore if a specific genetic change could lead to drug resistance by weakening the drug's attachment. Molecular dynamics (MD) simulations are a widely embraced tool for such biological inquiries. These simulations are rooted in Newton's classic equations of motion (F=ma), which describe how forces affect particles. In MD simulations, you begin with a known set of coordinates, like the arrangement of atoms in an enzyme-drug complex submerged in water. The method calculates forces acting on each atom, allowing you to track their movements over time based on Newton's equations.

The key is computing the forces by assessing the potential energy associated with atom positions, and this relies on the presence and positions of all other atoms. To do this practically, we employ a force field—a set of equations and parameters. MD simulations at the atomic scale use short time steps, typically 1-4 femtoseconds (fs, $10^{-15}$ seconds). So, even for a simulation lasting just 0.1 microseconds (μs), you'd perform a whopping 50-million-time steps. Figure 2 provides a visual depiction of this method.

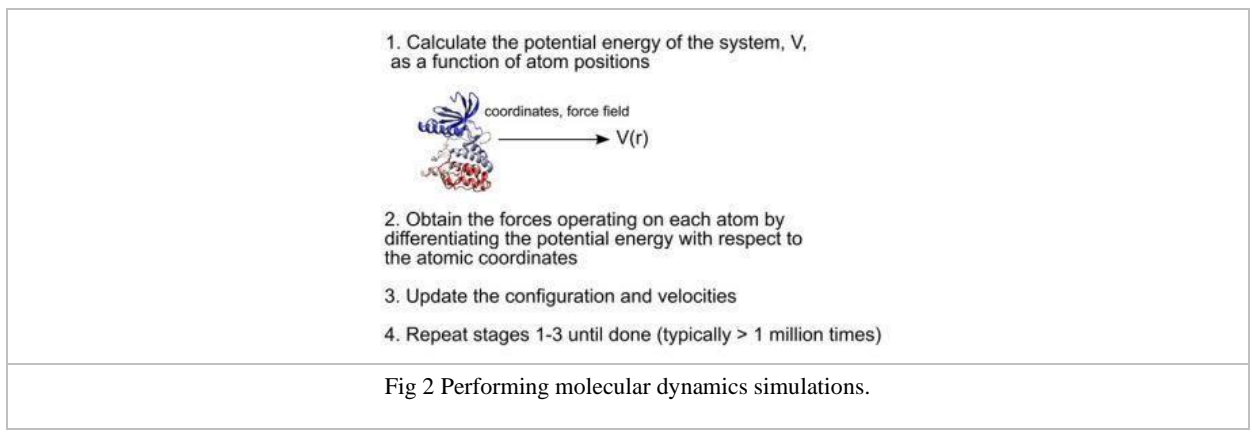

Fig 2 Performing molecular dynamics simulations.

After conducting an MD simulation, you obtain a dataset containing particle velocities and coordinates, commonly referred to as a trajectory. Various free and commercial software tools are accessible for visualizing these trajectory files. With these programs, researchers can essentially watch the simulation unfold, much like viewing a molecular movie. They have the flexibility to zoom in, zoom out, and focus on frames that seem particularly noteworthy or informative. This visualization capability enhances the researcher's ability to glean insights from the simulation data.

Comprehensively addressing what is MD in biological context Addressing will be extremely challenging, but recent examples shed light on the power of MD simulations. Transporter proteins, for instance, benefit greatly from MD simulations, as they provide insights into dynamic processes that structural and mutational data alone cannot answer. MD helps us understand how molecules are actually transported and the crucial roles of specific residues, even when their involvement isn't evident in known structural information.

In the realm of computer-aided drug design, MD simulations contribute by revealing both the correct and incorrect modes of molecule binding [28]. Furthermore, they elucidate the energetics governing binding reactions, such as the interplay between electrostatic and hydrophobic forces.

MD simulations are also instrumental in studying post-translational modifications and comparing modified and non-modified proteins. This comparative approach extends to understanding the mechanisms of action of drug molecules. Molecular dynamics (MD) simulations offer significant advantages but are not without limitations. Firstly, they excel in modeling soluble proteins with known structures that are unaltered and not bound to cofactors or inhibitors. Addressing systems involving multivalent metal ions, cofactors, membranes, or drug molecules often demand specialized protocols [29]. Secondly, MD simulations are constrained to time scales of less than 1 millisecond. Conformational changes at the level of whole protein domains require alternative methods or specialized hardware, which may introduce numerical inaccuracies [30, 31]. Thirdly, MD simulations employ an approximate potential energy function that may not fully capture certain system behaviors, although it generally performs well for biomolecules under standard conditions. Lastly, canonical MD simulations, based on classical mechanics, cannot effectively model reactions involving the formation or cleavage of covalent bonds. To overcome these limitations, alternative techniques such as quantum mechanics-based simulations (QM/MM and AIMD) and simplified models like coarse-grained MD have gained popularity, challenging the dominance of atomistic MD grounded in classical mechanics.

### 2.1.2. Brownian dynamics

#### 2.1.2.1. Advantages

Combining the principles of physics with lightning-fast usability, this approach excels at handling vast biomolecular complexes.

#### 2.1.2.2. Constrains

In the realm of molecular simulations, a common approach is to consider molecules as unyielding, rigid bodies. This means that in these simulations, no attention is paid to the intricate internal dynamics of the molecules themselves. Instead, they are treated as if they were immovable entities, a simplification that facilitates certain types of modeling and analysis within the context of molecular science.

#### 2.1.2.3. Practical use

Brownian dynamics proves invaluable when dealing with the interaction between two known protein structures and the limited data available about their complex formation, often obtained through techniques like cross-linking experiments. Brownian dynamics allows scientists to study complex formation, particularly when multiple binding modes may be in play.

In the field of biomedicine, Brownian dynamics (BD) simulations play a pivotal role in tracking the assembly of complex structures, particularly those involving multiple proteins. These systems are often too vast and progress too leisurely to be effectively scrutinized through direct molecular dynamics (MD) simulations. This leads to the necessity of employing approximate methods to decipher the intricate process of complex formation. Subsequently, MD simulations can be employed in a later phase to distinguish among various potential complex configurations.

In the realm of biomolecular Brownian dynamics (BD) simulations, the essence of force extraction lies within the surrounding electrostatic potentials of particles. BD simulations not only provide trajectories and encounter rates for these components but also offer a versatile tool for comparing diverse complexes. The applications of BD extend beyond the confines of protein-protein interactions. They find relevance in a spectrum of scenarios, including the simulation of ion migration in membrane channels [32], the investigation of enzyme-ligand interactions [33], and the construction of models for the intricate interplay between proteins and DNA [34], among other intriguing use cases.

### 2.1.3. Modelling based on quantum chemistry

#### 2.1.3.1. Advantages

Brownian dynamics simulations excel in accuracy and versatility. They can effectively model processes involving the formation and breaking of covalent bonds or explore the behavior of molecules in excited states, which are beyond the reach of less precise techniques.

#### 2.1.3.2. Constrains

Limitations of Brownian dynamics (BD) simulations include their slow execution, steep learning curve, applicability primarily to short timescales, and suitability for small-scale systems. 2.1.3.3. Practical use

Exploring the intricate interactions between carcinogens and nucleic acid bases is a complex endeavor. Quantum chemistry, grounded in the principles of quantum mechanics (QM), emerges as a powerful tool in this pursuit. Its distinctive advantage lies in its ability to provide superior accuracy when compared to methods like molecular dynamics, which rely on non-quantum energy functions. Quantum chemistry excels in handling processes involving bond-breaking and formation. However, it does come with some challenges. One notable drawback is the considerable computational cost associated with quantum chemistry programs. Additionally, comprehending the underlying quantum mechanical principles can be daunting for non-experts, compounded by the field's intricate jargon, making it less accessible in publications. Nevertheless, efforts are underway to mitigate these challenges and broaden quantum chemistry's applicability. Full-scale quantum mechanical calculations on large proteins remain rare due to computational constraints. However, emerging methods, such as those detailed in references [35, 36], are actively addressing this limitation. Instead, researchers often construct models of specific systems, like catalytic enzyme sites, focusing on a subset of directly involved atoms. The remainder of the system is either simplified or approximated using faster techniques like quantum mechanics/molecular mechanics (QM/MM). Density functional theory (DFT), pioneered by Hohenberg, Kohn, and Sham in the 1960s, has helped alleviate some of these traditional QM limitations. In the realm of cancer research, quantum chemistry plays a pivotal role. It unravels the intricacies of carcinogen-DNA interactions [37], contributes to chemotherapy design [38], elucidates histone deacetylation mechanisms [39], and finds applications in various other facets of biomedicine.

### 2.1.4. Ab initio molecular dynamics

#### 2.1.4.1. Advantages

Dynamic processes are accurately modeled using molecular dynamics simulations with the precision afforded by quantum chemistry.

### 2.1.4.2. constrains

This approach is constrained to very small systems, typically consisting of tens of atoms, and extremely short time scales on the order of $10^{-11}$ seconds.

### 2.1.4.3. Practical use

In the realm of enzymatic reactions, researchers employ a model system anchored in the active-site structure. Ab initio Molecular Dynamics (AIMD) methods stand at the forefront, seamlessly connecting molecular dynamics with quantum mechanics. In AIMD, Newtonian mechanics drive system simulations, while quantum-mechanical principles underpin atom-level force calculations. The method's precision is unquestionable, yet it remains less prevalent for studying biological macromolecules due to its computational demands and the need for specialized expertise. One notable application of ab initio molecular dynamics lies in cancer research, as exemplified by its use in radiotherapy investigations. Among various implementations of AIMD, Car–Parrinello Molecular Dynamics (CPMD) [40] enjoys popularity. Another approach, Path-Integral Molecular Dynamics [41], provides insights into phenomena like nuclear tunneling, observed in certain enzymes [42].

## 2.1.5. Quantum mechanics/molecular mechanics

### 2.1.5.1. Advantages

It makes it possible to apply quantum mechanics-based techniques to large atomic systems, such as proteins.

### 2.1.5.2. Constrains

It shares the limitation of being as computationally slow as quantum chemistry methods.

### 2.1.5.3. Practical use

In the context of enzymatic reactions, particularly when considering the involvement of all protein residues, Quantum Mechanics/Molecular Mechanics (QM/MM) methods emerge as a valuable tool. These techniques segment the system, with one part (typically the catalytic site of the enzyme and substrate) being analyzed using quantum mechanics, while the remainder is approximated using methods akin to standard molecular dynamics simulations. This unique approach empowers researchers to delve into the intricacies of macromolecules, even when chemical bonds are formed and broken during reactions. The computational efficiency of QM/MM is closely tied to the speed of quantum mechanical calculations for the specific region of interest. However, it's worth noting that effectively implementing QM/MM requires a deep understanding of both quantum mechanics and molecular mechanics, as well as their harmonious integration. Nevertheless, QM/MM-based methods are widely adopted in the field of biology, offering applications with significant relevance to cancer research. Examples include modeling drugs that form covalent bonds with DNA [43] and unraveling the mechanism of mitogen-activated protein kinase [44].

2.1.6. Normal mode analysis

        2.1.6.1. Advantages

Streamlined, swift, and reasonably straightforward to execute — offering a rapid means to gain insight into atomic-level dynamics without the need for extensive simulations.

        2.1.6.2. Constrains

Lacks correlation with specific timescales and doesn't possess the breadth and precision of Molecular Dynamics (MD).

        2.1.6.3. Practical use

In the study of complex proteins comprising multiple domains, understanding their inter-domain movements is crucial for their functional relevance. Normal mode analysis (NMA) assumes equilibrium conditions while modeling protein motions as harmonic oscillations, distinguishing between low-frequency collective motions (e.g., domain interactions) and high-frequency local fluctuations. NMA offers simplicity and speed compared to molecular dynamics (MD) simulations, instantly separating global physiological motions from local deformations. However, there are limitations: proteins' motions aren't strictly harmonic, and extensive preparation is needed to bring the system close to its energy minimum. Despite these drawbacks, NMA can employ simplified models, making it computationally feasible for protein studies [45].

# Fundamentals of Molecular Dynamics Simulations

The fundamental concept underlying molecular dynamics (MD) simulations is quite intuitive. When you have detailed information about the positions of all the atoms within a complex biological system, like a protein immersed in a watery environment or even surrounded by a lipid bilayer, it becomes possible to compute the interactions and forces acting on each atom. This information is then utilized to apply Newton's laws of motion, enabling us to forecast the evolving spatial coordinates of every atom over time. Essentially, this involves progressing through time steps, iteratively determining the forces experienced by each atom, and subsequently adjusting their positions and velocities based on these calculated forces. The outcome of this process essentially creates a three-dimensional cinematic representation, akin to a movie, that chronicles the atomic-level arrangement of the system at every instance throughout the simulated time span.

These simulations possess formidable capabilities for several compelling reasons. Firstly, they provide an exhaustive account of the spatial positions and movements of each atom at every time step, a feat challenging to achieve through experimental methods. Secondly, these simulations afford precise and controlled conditions, encompassing factors such as the initial conformation of a protein, the presence of bound ligands, potential mutations or post-translational modifications, the composition of the surrounding molecular environment, protonation states, temperature, membrane voltage, and more. By comparing simulations conducted under varying conditions, it becomes possible to pinpoint the impacts of diverse molecular alterations and disturbances.

In an MD simulation, the determination of forces relies on a model known as a molecular mechanics force field. This force field is constructed by fitting it to the outcomes of quantum mechanical computations and often incorporates certain experimental data. To illustrate, a typical force field encompasses terms that account for electrostatic (Coulombic) interactions among atoms, spring-like components that emulate the optimal length of each covalent bond, and terms that encapsulate various other forms of interatomic interactions. It's important to note that these force fields are inherently approximations. While recent years have witnessed significant enhancements in force field accuracy through comparisons with diverse experimental datasets [46], they are not flawless, and the inherent uncertainties stemming from these approximations must be taken into consideration when interpreting simulation outcomes.

Furthermore, it's essential to understand that in classical MD simulations, no covalent bonds are formed or broken. To investigate reactions involving alterations in covalent bonds or those induced by light absorption, researchers frequently employ quantum mechanics/molecular mechanics (QM/MM) simulations. In QM/MM simulations, a small portion of the system is described using quantum mechanical calculations, while the remainder is simulated using MD techniques [47]. This approach is particularly valuable for examining reactions that entail changes in covalent bonds or are driven by photon absorption.

To maintain numerical stability, MD simulations require extremely short time steps, typically on the order of just a few femtoseconds (10e-15 seconds) each. Many of the biologically significant occurrences, such as critical structural transformations in proteins, occur over much longer timescales, ranging from nanoseconds to microseconds or even longer. Consequently, a standard simulation involves the execution of millions or even billions of these tiny time steps.

Adding to the computational intensity, each of these individual time steps necessitates the evaluation of millions of interatomic interactions. This high computational demand is a hallmark of MD simulations, making them resource-intensive and requiring advanced computing infrastructure to handle the complex calculations effectively.

In the course of the past few decades, significant strides in computing technology, as well as advancements in the algorithms and software underpinning MD simulations, have ushered in an era of longer and more cost-effective simulations. The recent progress in this field has been nothing short of remarkable. Specialized hardware, as highlighted by Shaw and colleagues [48-50], has played a pivotal role in substantially boosting the maximum achievable simulation speed. This breakthrough has enabled certain simulations to extend their reach into timescales measured in milliseconds.

Even more significantly, the advent of Graphics Processing Units (GPUs) has revolutionized MD simulations. These GPUs have empowered simulations running on just one or two affordable computer chips to outperform what was once only attainable on some of the most powerful supercomputers, as observed by Salomon-Ferrer and collaborators [50]. This democratization of computing power through GPUs has made biologically relevant timescale simulations accessible to a far broader spectrum of researchers than ever before.

It's true that conducting simulations has become comparatively more straightforward nowadays, as discussed in the section on "Practical Considerations in Using MD Simulations." Additionally, computational resources required to perform substantial amounts of simulation work have become increasingly accessible to a wider audience. However, the realm of expertise lies in identifying which questions can effectively be tackled through simulations, crafting simulations tailored to address these specific inquiries, and subsequently deriving meaningful insights from the wealth of trajectory data that encapsulates the dynamic behavior of numerous atoms—a task that can prove particularly challenging. Furthermore, there exists a plethora of advanced simulation techniques designed to address questions that cannot be tackled through straightforward "brute force" simulations. These advanced methods often necessitate a deep understanding of both the biological questions at hand and the intricacies of simulation methodology.

# Atomic Force Field Model of Molecular Systems

The atomic force field model characterizes physical systems as assemblies of atoms held together by interatomic forces, where chemical bonds arise from the shape of atom-to-atom interactions within molecules. This interaction is defined by the potential energy function $U(r_1,\ldots,r_N)$, which quantifies the energy of N interacting atoms based on their positions, $r_i = (x_i, y_i, z_i)$. The force experienced by each atom is determined by the gradient, as depicted in equation (1).

$$F_i = -\nabla_{r_i} U(r_i, \ldots, r_N) = -(\partial U \partial x_i, \partial \partial U y_i, \partial \partial U z_i) \qquad (1)$$

The concept of 'atoms in molecules' is a simplified representation of the quantum-mechanical reality, where molecules consist of interacting electrons and nuclei. Electrons exhibit some degree of delocalization, shared by multiple nuclei, shaping chemical bonds. The adiabatic (or Born-Oppenheimer) approximation, grounded in the significant mass disparity between nuclei and electrons, allows the separation of electronic and nuclear concerns with high accuracy.

The electron cloud rapidly achieves equilibrium for each fleeting but quasi-static heavy nuclei configuration. Simultaneously, the heavy nuclei maneuver within the domain defined by the averaged electron densities. Consequently, we can introduce the idea of a potential energy surface, which dictates the nuclei's dynamics without requiring explicit consideration of the electron behavior. With this potential energy surface in hand, we can employ classical mechanics to trace the trajectory of the nuclei.

When we pinpoint the nuclei as the core of atomic entities and delineate the adiabatic potential energy landscape through an implicit interaction framework, we establish a robust validation for the intuitive portrayal of a molecule as an assembly of interacting atoms. This demarcation of electronic and nuclear aspects also signifies that instead of tackling the formidable quantum electronic dilemma, which can often be impractical, we can adopt an alternative approach. In this alternative strategy, we encapsulate the influence of electrons on the nuclei through the use of an empirical potential function.

The challenge of discovering a plausible potential capable of faithfully emulating the genuine energy landscapes presents a formidable endeavor. However, this pursuit yields substantial simplifications in computational tasks. Atomic force field models and classical molecular dynamics draw upon empirical potentials characterized by specific functional formulations that encapsulate the underlying physics and chemistry of the targeted systems. These potentials are calibrated by selecting adjustable parameters that ensure a close alignment with pertinent segments of the ab initio Born-Oppenheimer surface. Alternatively, they may find their foundation in empirical data. In simulations concerning biological systems, a typical force field conforms to the pattern exemplified in equation (2).

$$U(r_1,\ldots,r_N) = \sum_{bonds} a_{2i}(l_i - l_{i0})2 + \sum_{angles} \frac{b_i}{2}(\theta_i - \theta_{i0})^2 + \sum_{torsions} c_{2i}[(1 + \cos(n\omega_i - \gamma_i)] +$$

$$\sum_{atoms\ pairs} 4\varepsilon_{ij}\left[\left(\frac{\sigma_{ij}}{r_{ij}}\right)^{12} - \left(\frac{\sigma_{ij}}{r_{ij}}\right)^6\right] + \sum_{atoms\ pairs} k\ q^i r_{ij} q^j$$

(2)

In the initial three components of the equation, the summation variables encompass the entirety of bonds, angles, and torsion angles as delineated by the covalent configuration of the system. Conversely, within the latter two elements, the summation variables span across all possible atom pairs (or locations hosting point charges represented as $q_i$) that are separated by distances $r_{ij}$ equal to the absolute difference between the positions of $r_i$ and $r_j$, without any chemical bonding connection.

In the realm of physical descriptions, the initial two terms encapsulate the energies associated with deviations in bond lengths (denoted as $l_i$) and bond angles ($\theta_i$) from their respective equilibrium values, represented as $l_{i0}$ and $\theta_{i0}$. These terms adopt a harmonic form, governed by force constants ($a_i$ and $b_i$), which preserves the chemical structure accurately. However, it confines the model from representing chemical transformations like bond breaking. The third term addresses rotations around chemical bonds, featuring periodic energy variations with a periodicity parameter ($n$) and energy barriers determined by $c_i$. The fourth term characterizes interatomic forces, both repulsive and attractive (dispersive), in the form of the Lennard-Jones 12-6 potential. Lastly, the fifth term pertains to the Coulomb electrostatic potential.

Specific environmental effects are accommodated through adjusted partial charges ($q_i$) and a constant ($k$), along with adaptable van der Waals parameters ($\varepsilon_{ij}$ and $\sigma_{ij}$).

# Molecular Dynamics Algorithm

In the context of MD simulations, the objective lies in observing the unfolding temporal evolution of a cluster of interacting particles. This is achieved through the application of Newton's equations of motion, as denoted by equation (3). These equations function as the guiding framework, shedding light on the intricate interplay among particles and their progression over time.

$$F_i = m_i \frac{d^2 r_i(t)}{dt^2} \quad (3)$$

In this framework, the variable $r_i(t)$ denotes the position vector of the $ith$ particle, encapsulating its spatial coordinates $(x_i(t), y_i(t), z_i(t))$.. Simultaneously, $F_i$ represents the force exerted upon the $ith$ particle at the specific instant in time denoted as t, while mi stands as the mass characteristic to the particle.

'Particles' typically refer to atoms, though they can encompass various distinct entities, such as specific chemical entities, that can be conveniently characterized using a specific interaction law. In order to tackle the integration of the second-order differential equations mentioned above, one must define the instantaneous forces acting on these particles, along with their initial positions and velocities. Given the inherent complexity of many-body systems, it becomes necessary to discretize the equations of motion and resolve them through numerical methods. These molecular dynamics (MD) trajectories are defined by vectors representing both position and velocity, providing a comprehensive depiction of the system's evolution in phase space. Consequently, the positions and velocities undergo propagation with finite time intervals through the application of numerical integration techniques, such as the Verlet algorithm. The time-varying positions of particles within three-dimensional space are formally characterized as $r_i(t)$, whereas the associated velocities, $v_i(t)$, play a pivotal role in determining the kinetic energy and temperature of the system under investigation.

# Numerical Integration of the Equations of Motion

The primary goal of numerically integrating Newton's equations of motion is to derive an expression that characterizes the positions, denoted as $r_i(t + \Delta t)$, at the time $t + \Delta t$, based on the known positions at the current time, $t$. In the realm of molecular dynamics MD simulations, the Verlet algorithm stands out as a popular choice due to its simplicity and inherent stability. The fundamental equation for this algorithm can be obtained through Taylor series expansions applied to the positions $r_i(t)$, and it is articulated as indicated in equation (4).

$$r_i(t + \Delta t) \cong 2r_i(t) - r_i(t - \Delta t) + \frac{F_i(t)}{m_i}\Delta t^2 \quad (4)$$

Equation (4) retains its accuracy up to terms of the fourth power in $\Delta t$. When it comes to calculating velocities, there are two viable approaches: one can compute them from the positions, or they can be explicitly propagated, as seen in alternative schemes such as the leapfrog or velocity Verlet methods.

In the trajectory calculations, achieving exact trajectories necessitates an infinitesimally small integration step. However, for practical purposes, it becomes imperative to employ larger time steps to effectively sample longer trajectories. The selection of an appropriate time step, $\Delta t$, is a critical consideration, primarily influenced by the swiftness of motions within the system. A pertinent example lies in the dynamics of chemical bonds, especially those involving lightweight atoms like the O-H bond, which undergo rapid vibrations with periods spanning several femtoseconds. To uphold the stability of the integration process, it becomes essential for Dt to operate at a subfemtosecond scale. While constraints can be introduced within the integration algorithm to suppress the influence of the fastest but non-essential vibrations, the attainment of a time step exceeding 5 femtoseconds remains an elusive goal in the simulation of biomolecules. This practical limitation underscores the intricate balance between capturing accuracy and computational feasibility in molecular dynamics simulations.

## Insights Unveiled by MD Simulations

Molecular dynamics (MD) simulations offer a versatile tool for addressing a wide array of questions, as depicted in Figure 2. In this discussion, we'll explore some of the most prevalent use cases, with a particular focus on how simulations often enhance and complement experimental inquiries within the field of molecular biology. For more concrete examples, Figures 3, 4, and 5 provide visual representations of select studies we've conducted based on simulations.

Simulation stands as a versatile tool, offering a fundamental and intuitive means of exploring the mobility and adaptability within biomolecules. Traditional techniques like X-ray crystallography and cryo-EM provide average structural insights. However, through simulation, we gain the ability to quantitatively assess how different regions of the molecule move under equilibrium conditions and the nature of their structural fluctuations. Furthermore, these simulations unveil the dynamic interplay of vital components, such as water molecules and salt ions, which often exert a critical influence on protein functionality and the binding of ligands [51, 52].

Simulation serves a multifaceted role, including the validation and enhancement of structural models. For instance, crystallographic structures may exhibit anomalies due to the inherent packing of the crystal lattice, or in the case of membrane proteins, the absence of a lipid bilayer can introduce inaccuracies. These issues can often be mitigated by initiating simulations based on the crystal structure within an appropriate solvent environment, allowing the structure to adapt to a more favorable conformation, if possible [53].

Likewise, simulations play a critical role in validating the binding orientations of ligands. A binding pose that remains stable throughout a simulation is considered more reliable than an unstable one [54]. This approach has been particularly valuable in resolving ligand positions within cryo-EM structures characterized by uncertain ligand density [55]. It's important to note that while molecular dynamics simulations can be effective in refining protein homology models, their success rate can vary, with some attempts yielding positive results while others prove less fruitful [56, 57].

In contrast, MD simulations serve as versatile tools in the realm of structural modeling, either in constructing or refining models based on empirical structural data from the field of biology. Consider, for instance, the X-ray crystal structures. Here, an ingenious MD-driven simulated annealing technique steps in, seamlessly melding the model with experimental data, all the while preserving the essential physical integrity [58, 59]. This innovative method excels in rectifying model imperfections that defy conventional least-squares regression techniques.

Moreover, the application of MD-based protocols extends its prowess to the realm of molecular architecture. In the case of low-resolution cryo-EM density maps, it emerges as the go-to method, particularly when high-resolution blueprints for individual components within a complex are independently accessible [60, 61]. But the utility doesn't end there. MD simulations shine in the arena of data interpretation, effortlessly extracting ensembles of conformations—rather than just a solitary structure— from NMR data [62]. In all these instances, the molecular mechanics force field receives a judicious infusion of terms rooted in experimental data, culminating in structures, or ensembles thereof, that bear an uncanny resemblance to the empirical evidence, all while minimizing energy levels.

An essential application of MD simulation involves studying how a biomolecular system responds to various perturbations. These perturbations encompass a wide range of possibilities. For instance, one can begin by removing a ligand that is bound to an experimentally determined protein structure. Subsequent simulations reveal the profound impact of ligand removal on the protein's conformation ([63-65]; Figure 3). Alternatively, the simulation can involve the replacement of an existing ligand with a different one or the introduction of a ligand in a location where the experimental structure lacked one [66, 67]. Another avenue explores mutating one or more amino acid residues within the protein. This can be employed to understand the functional consequences of mutations or to reconcile experimentally resolved constructs with the wild-type structure in cases where they differ ([68]). Further possibilities include simulating posttranslational modifications, such as phosphorylation or other chemical alterations [69, 70], modifying the protonation state of acidic or basic amino acids [71], applying external forces to simulated atoms to mimic transmembrane voltage or mechanical strain [72], or altering the molecular environment of a simulated protein, including changes in salt concentration or lipid composition within a membrane. In all these cases, it is generally advisable to perform multiple simulations of both perturbed and unperturbed systems to discern consistent differences in the results.

Many MD simulations serve the purpose of observing dynamic biomolecular processes, particularly crucial functional phenomena like ligand binding, ligand- or voltage-induced structural changes, protein folding, or membrane transport. These simulations offer a unique avenue for addressing structural questions that may be challenging to investigate experimentally. For instance, MD simulations can shed light on the intricate process of protein folding, revealing the sequential formation of substructures [73, 74]. They can elucidate how the binding of a ligand to a GPCR's extracellular surface triggers consequential alterations on the intracellular side where the G protein binds [75]. Furthermore, MD simulations can provide valuable insights into the structural underpinnings of protein allostery, a fundamental phenomenon in proteins ([76]; Figure 5). 2Additionally, these simulations can investigate the mechanics of alternating access transporters, elucidating how they prevent both their outer and inner gates from opening simultaneously ([77-79]; Figure 4). They can delve into the factors that govern the kinetics of ligand binding and dissociation ([65, 80, 81]. Moreover, MD simulations can explore the structural intricacies behind the transport of water and ions across biological membranes [82-85]. Intrinsically disordered proteins, a fascinating area of study, can also be investigated through MD simulations. They offer a window into the assembly process of these proteins as they come together to form fibrils [86, 87]. In certain instances, a solitary, unguided simulation has the remarkable capacity to encompass an entire complex process. However, there are scenarios where this approach may prove challenging. This could be due to exceedingly protracted timescales or the involvement of reactive chemistry. Nevertheless, even in such intricate cases, it is often feasible to reconstruct the process through alternative means. For instance, one can opt to simulate distinct segments of the process separately. Alternatively, a repertoire of enhanced sampling simulation methods can be brought into play. These techniques, exemplified by the work of Bernardi et al. [88], Harpole et al. [89], Hertig et al. [76], and Schwantes et al. [90], offer invaluable strategies for dissecting and comprehending intricate processes that elude direct, uninterrupted simulation.

# Literature Review

Timpson et al. [91] discusses the importance of understanding and targeting the metastatic process in cancer research. Despite advancements in understanding primary cancer, metastasis remains the leading cause of cancer-associated deaths. Researchers aim to identify key steps in metastasis that can be targeted for therapeutic interventions to improve treatment strategies. The study emphasizes the significance of intravital imaging techniques in gaining insights into the dynamic molecular dynamics of cancer cell invasion and metastasis. By utilizing real-time imaging, researchers can visualize and analyze dynamic biomarkers within living animals. The use of photobleaching and photoactivation techniques enables the assessment of therapeutic interventions during the early stages of tumor cell metastasis. The paper concludes by highlighting the value of quantitative real-time imaging in designing stage-specific drug regimes for the treatment of metastatic cancer. By elucidating the susceptible steps in the metastatic process, researchers can develop targeted therapeutic interventions to combat cancer spread and enhance patient outcomes.

Tripathi et al. in their research [92] focused on investigating the molecular dynamics and free energy landscape methods to understand the resistance mechanisms caused by specific leucine point mutations (L215H, L217R, and L225M) in βI-tubulin, resulting in paclitaxel (Ptxl) resistance in cancer cells. Using computational techniques such as molecular docking, molecular dynamics simulation, and energy estimation, the study reveals that the mutations reduce the binding affinity of Ptxl and disrupt its interactions with key regions like the M-loop, S6-S7 loop, and H6-H7 loop. The flexibility of the M-loop region is found to be crucial for microtubule stabilization, and its impaired interaction with Ptxl in mutant types contributes to resistance. This research provides insights into the molecular basis of Ptxl resistance, aiding in the development of new microtubule stabilizers targeting the mutant types. The output of this research was a comprehensive understanding of the molecular mechanisms behind paclitaxel (Ptxl) resistance in cancer cells caused by specific leucine point mutations in βI-tubulin. Through the use of computational methods and analysis, the study revealed that these mutations reduce the binding affinity of Ptxl and disrupt its interactions with critical regions involved in microtubule stabilization. The research highlighted the importance of the M-loop flexible region's interaction with Ptxl for microtubule stability and provided valuable insights into the resistance mechanisms. This knowledge can be used to develop novel microtubule stabilizer molecules that are effective against the mutant types, potentially leading to improved treatment strategies for Ptxl-resistant cancers.

The study by Kim et al.[93] focuses on the therapeutic potential of embelin, a quinone derivative found in the fruits of Embelia ribes Burm. Embelin has been recognized for its various beneficial properties, including anti-tumor, anti-inflammatory, and anti-bacterial activities. Inflammation, regulated by the cytokine tumor necrosis factor alpha (TNF-α), plays a significant role in several human pathologies, such as cancer and neurodegeneration. The research investigates the impact of embelin on TNF-α production and its associated anti-inflammatory activity. Specifically, the study explores the inhibition of TNF-α converting enzyme (TACE), responsible for releasing the soluble component of pro-TNF-α into the extracellular space. TACE inhibition has been proposed as a potential therapeutic strategy for inflammation and cancer.

Combining molecular dynamics simulations and experimental evidence, the study provides insights into embelin's ability to inhibit TACE and reduce TNF-α production. These findings highlight the potential of embelin as a therapeutic agent for targeting inflammation and cancer cell metastasis. The research contributes to the understanding of the molecular mechanisms underlying embelin's anti-inflammatory effects and supports its potential as a novel therapeutic approach.

Tang et al. in [94] focused on the development of biodegradable nano prodrugs for cancer therapy. These nano prodrugs aim to address the limitations of chemotherapy drugs, such as side effects on normal tissues.

In this study, transferrin modified $MgO_2$ nanosheets are designed as biodegradable nano prodrugs to selectively deliver reactive oxygen species (ROS) to cancer cells for molecular dynamic therapy. The nanosheets are designed to take advantage of the tumor microenvironment, which is acidic and has low catalase activity. Within this environment, the nanosheets react with protons to release nontoxic $Mg^{2+}$ ions. Simultaneously, this reaction generates abundant hydrogen peroxide ($H_2O_2$), which induces cell death. Furthermore, the structure of transferrin is damaged, leading to the release of $Fe^{3+}$ ions. The $Fe^{3+}$ ions react with the $H_2O_2$ to produce highly toxic hydroxyl radicals (·OH), which further contribute to killing tumor cells. By utilizing these biodegradable nano prodrugs, the researchers demonstrate a strategy for selectively delivering ROS to cancer cells, enabling effective molecular dynamic therapy. This approach holds promise as a next-generation cancer therapy, harnessing the advantages of nanotechnology and addressing the limitations of traditional chemotherapy drugs. The study provides insights into the design and functionality of biodegradable nano prodrugs for targeted cancer treatment and highlights their potential in improving therapeutic outcomes.

In their study [95] Raffaini et al. investigated the potential use of γ-cyclodextrins (γ-CD) complexed with C70 fullerenes as photosensitizers for cancer therapy. They utilized molecular dynamics methods, including molecular mechanics and simulations, to analyze the stability and geometry of the 2:1 complexes [(γCD)2/C70]. Comparisons were made with C60 complexes, which were found to be less efficient in cancer therapy. The researchers observed that the γ-CD/C70 complex had a different arrangement and lower stability compared to C60 complexes. In water, the complex formed relatively stable arrangements while exposing part of C70 to the solvent. Understanding the behavior of these complexes through molecular dynamics simulations contributes to enhancing solubilization and facilitating fullerene insertion into liposomes or cell membranes, with implications for improving cancer therapy using photodynamic treatment.

In the research conducted by Martínez-Muñoz et al. [96], the focus was on developing new therapeutic strategies against breast cancer by targeting the G Protein-Coupled Estrogen Receptor 1 (GPER1). The authors employed ligand-based virtual screening and molecular dynamics simulations to investigate the binding mechanism of a set of G15/G1 analogue compounds. They designed a novel molecule called G1PABA, which exhibited promising binding characteristics with GPER1. Experimental studies confirmed the inhibitory effects of G1-PABA on cell proliferation in breast cancer cell lines (MCF-7 and MDA-MB231) and a normal cell line (MCF-10A). The compound demonstrated concentration-dependent inhibition of cell proliferation, with better efficacy observed in the absence of phenol red, which was found to interfere with G1-PABA's action on GPER1. These findings highlight the potential of G1-PABA as a therapeutic agent targeting GPER1 in breast cancer treatment.

In the study by Gade and colleagues [97] focus was on elucidating the chemo sensitization potential of acridones, which have shown promise in reversing multidrug resistance (MDR) in cancer treatment. The researchers employed various molecular modeling techniques to investigate the interaction between acridones and P-glycoprotein (P-gp), an important factor in MDR. Pharmacophore modeling was performed to identify key features of chemo sensitizing acridones, resulting in the identification of an efficient pharmacophore containing two hydrogen bond acceptors and three aromatic rings (AARRR.14). The dataset was further screened against this pharmacophore, leading to the identification of 25 best-fit molecules from the NCI 2012 chemical database. Gaussian-based QSAR studies were performed to predict the favored and disfavored regions of the acridone molecules. Atom-based QSAR analysis yielded regression analysis results with r2 of 0.95 and q2 of 0.72, while field-based QSAR analysis showed r2 of 0.92, q2 of 0.87, and r2cv of 0.71. These analyses provided insights into the potential regions of the compound and its preferred interactions. Molecular dynamics simulations were performed for compound 10 and human P-gp, obtained from homology modeling. These simulations allowed for the analysis of conformational changes occurring during the interaction of compound 10 with P-gp, providing insights into the fate of the acridone molecule in the P-gp environment. By combining the data from different in silico techniques, the researchers gained a deeper understanding of the structural and mechanistic aspects of acridone interaction with P-gp. This study serves as a strategic basis for designing more potent molecules with anti-cancer and multidrug resistance reversal activities.

The paper by Wang and colleagues [98] investigates how trehalose lipids selectively attack cancer cells using molecular dynamics simulations. The study finds that a specific lipid composition strongly interacts with cancer cell membranes but not with normal cells. This interaction leads to a potent anti-cancer effect without harming normal cells. The simulations reveal that certain lipids with hydrophilic and bulky head groups accumulate, causing the liposomes and cancer cell membranes to closely interact. In contrast, no significant interaction occurs between the liposomes and normal cell membranes. The study's findings reveal that a specific composition of trehalose lipids (TreC14) causes a strong interaction with cancer cell membranes but not with normal cells. This interaction leads to the bending and contact of the liposomes and cancer cell membranes. Lipids with hydrophilic and bulky head groups, including TreC14, phosphatidylinositol (PI), and phosphatidylserine (PS), accumulate and form domains, resulting in positive curvatures and tight contact between the liposomes and cancer cell membranes. This interaction is not observed with normal cell membranes due to the absence of PI and PS. The degree of membrane distortion correlates with the inhibitory effect on cancer cell growth, suggesting a potential mechanism for the anticancer activity of DMTreC14 liposomes. However, complete fusion of the liposomes and cancer cell membranes was not observed in the simulations, indicating the need for further investigation. The differences in lipid distributions between cancer and normal cells contribute to the selective interaction of DMTreC14 with cancer cells, offering insights into the potential efficacy of DMTreC14 liposomes as cancer treatment agents.

Omidvarborna et al. [99] investigate the molecular mechanism of the reaction between cold atmospheric pressure plasma (CAP) and cancer cells at the atomic level. Specifically, it focuses on the damage caused to DNA molecules by three reactive oxygen species (ROS): O, OH, and H2O2. The study utilizes a reactive force field to simulate the interaction between CAP ROS and DNA. The simulation reveals that the ROS react with DNA in cancer cells, leading to the disruption of important chemical bonds within DNA molecules and causing damage to cancer cell genes. Consequently, this reaction affects gene transcription, replication, and reproduction in cancer cells, ultimately inhibiting their unlimited proliferation. The study highlights that different ROS species can inflict varying degrees of damage to DNA molecules at distinct structural positions, resulting in significant and irreversible molecular structure damage to the DNA. The study used simulations to examine the interaction between reactive oxygen species (ROS) and DNA molecules. The results showed that ROS in plasma can damage DNA by breaking important chemical bonds. This damage affects cell genes and can induce apoptosis. Different types of ROS cause DNA damage at different positions, leading to irreversible structural changes. The simulations revealed hydrogen capture reactions between ROS and DNA, with O atoms being more reactive than OH radicals. The efficiency of subsequent reactions was higher for OH radicals. The ratio of broken N-H bonds varied among different base pairs. Overall, the findings shed light on how ROS damage DNA and contribute to cancer cell apoptosis.

Zaccai in [100] examines the application of incoherent neutron scattering as a method for investigating molecular dynamics within cells. It reviews experiments conducted to study intracellular molecular dynamics in various organisms, such as bacteria, archaea, red blood cells, brain cells, and cancer cells. The research primarily focuses on understanding water diffusion within cells and the impact of proteome molecular dynamics on cellular adaptation to temperature, pressure, and salt stress. Additionally, the paper briefly discusses the potential correlations between neutron scattering findings and molecular dynamics simulations in the study of intracellular dynamics. The simulation advances in silico whole-cell models, connecting molecular experiments in cell biology. Neutron scattering results can help validate simulations.

Understanding the behavior and characteristics of cancer cells is crucial for early cancer diagnosis and treatment. The study by Kashab et al. [101] utilizes molecular dynamics to investigate the atomic behavior and stability of cancer cells under various external forces and pressures. Parameters like gyration radius, interaction energy, and force are examined. The results reveal that increasing external force and pressure impact the gyration radius, interaction energy, and force values. The findings emphasize the significance of comprehending atomic behavior in disease diagnosis and treatment, particularly for cancer.

Nguyen et al. [102] investigate the mechanical differences between normal cells and cancer cells, focusing on alterations in the cell membrane. Molecular dynamics simulations are conducted on normal and cancer cell membranes to compare their structural and elasticity properties. The overexpression of phosphatidylserine lipids does not significantly affect the cancer membranes, but a reduction in cholesterol concentration leads to noticeable changes in membrane properties, particularly decreases in bending, tilt, and twist moduli. This suggests that lower cholesterol levels in cancer membranes may contribute to the softening of cancer cells.

Kordzadeh et al. [103] explore the interaction between functionalized carbon nanotubes (f-CNT) and cancerous cell membranes in the context of doxorubicin drug delivery. Molecular dynamics (MD) simulations and Density Functional Theory (DFT) methods are utilized to investigate the process of drug loading onto f-CNTs. The findings indicate that the interaction between doxorubicin molecules and f-CNTs is stronger in the case of f-CNTs functionalized with carboxyl and folic acid (CNT-COO-FA) compared to those functionalized with carboxyl only (CNT-COO). Simulations of drug release near a lipid bilayer, representing cancerous cells, demonstrate that CNT-COO-FA exhibits a pH and ligand-sensitive mechanism, resulting in enhanced drug release efficiency. Conformational changes in the lipid bilayer and folate receptor during drug release are evaluated, revealing significant alterations in the presence of CNTCOO-FA. The results highlight the superior performance of CNT-COO-FA in delivering doxorubicin, attributed to both pH and ligand sensitivity, while CNT-COO primarily enhances drug delivery through pH sensitivity.

Rivel et al. [104] focused on developing realistic membrane models for normal and cancer cells, specifically examining their cholesterol content. The results demonstrate that the loss of lipid asymmetry in cancer cell membranes significantly reduces the permeability of cisplatin by approximately one order of magnitude compared to normal cell membranes. Additionally, changing the cholesterol molar ratio from 0% to 33% also leads to a similar one-order-of-magnitude decrease in membrane permeability. Moreover, it is noted that pure DOPC membrane, commonly used in drug permeability studies, exhibits permeability rates that are 5-6 orders of magnitude higher than membranes with realistic lipid compositions, suggesting its inadequacy as a model in studying drug permeability.

Tabrez et al. [105] focused on targeting glutaminase (GLS) as a potential strategy to suppress cancer progression. By employing a sequential screening approach using a traditional Chinese medicine (TCM) database, followed by drug-likeness assessment and molecular dynamics simulations, 12 potent compounds were identified as potential GLS inhibitors. Among them, ZINC03978829 and ZINC32296657 showed higher binding energy values compared to the control compound, indicating their potential efficacy in inhibiting GLS. Molecular dynamics simulations further confirmed the stable interaction between these compounds and GLS, with the formation of hydrogen bond interactions. While these findings suggest the potential of these compounds as GLS inhibitors, further laboratory testing is necessary to optimize their effectiveness in cancer management.

Esophageal adenocarcinoma (EAC) is a rapidly growing cancer associated with chronic gastroesophageal reflux disease (GERD). The exposure of the esophageal epithelium to stomach acid during GERD triggers gene mutations, potentially leading to EAC development. While p53 is activated to eliminate mutated cells, NFKB coordinates the healing of remaining cells. However, if TP53 mutations occur (which is common), the mutant product promotes tumorigenesis, and NFKB collaborates with mutant p53 to drive cancer progression. Although TRAIL, a cytokine produced during GERD, can selectively kill cancer cells, its clinical effectiveness is limited due to the development of defense mechanisms by cancer cells. To overcome this obstacle, a second agent is needed to disarm the cancer cells. CCN1 emerges as a promising molecule that promotes normal esophageal cell growth, inhibits NFKB, suppresses malignant transformation, and induces EAC cell death through TRAIL-mediated apoptosis. The article by T. Dang and J. Chai [106] sheds light on the molecular dynamics underlying EAC and suggests potential strategies for targeting the disease.

Ihmaid et al. [107] focused on the discovery of novel compounds with selective anti-breast cancer activity and CDK9 inhibition. Triaromatic flexible agents bearing a 1,2,3-triazole ring and terminal lipophilic fragments were synthesized. The compounds showed significant anticancer activity against aggressive breast cancer cell lines with micromolar IC50 values. CDK assays and molecular docking simulations supported their mechanism of action. Compound 34 exhibited high selective cytotoxicity against breast cancer cells. Molecular dynamics simulations confirmed its stable interaction with CDK9. Compound 34 shows promise for further drug design optimization in breast cancer treatment.

In this study [108] molecular dynamics simulations were used to explore the interaction between a drug carrier, TiO2 nanosheet, and an RNA aptamer. Various parameters, including receptor-aptamer interaction, structural dynamics, and the influence of water molecules, were analyzed. The results showed that the aptamer can interact with the receptor, both with and without the nanosheet. Notably, a stronger interaction was observed when the aptamer detached from the nanosheet. The research showed that the RNA aptamer showed interaction with the cancer cell receptor, both in the presence and absence of the TiO2 nanosheet. The strongest interaction occurred when the aptamer detached from the nanosheet.

The study by Parthiban et al. [109] focused on the isolation and characterization of 7-hydroxy flavone from Avicennia officinalis L. leaves. The compound exhibited promising anticancer activity against cervical and breast cancer cell lines. Additionally, it demonstrated antioxidant properties. Molecular docking and molecular dynamics simulations revealed its strong binding affinity with the anti-apoptotic Bcl-2 protein. These findings suggest the potential of 7-hydroxy flavone as a drug candidate for cancer treatment. The study provides valuable insights into the biological evaluation and molecular interactions of this compound, paving the way for further research and development in the field.

Timpson et al. [110] presents a comprehensive overview of imaging molecular dynamics in living organisms, specifically focusing on cancer-related studies in mice. The authors highlight the significance of the microenvironment in shaping developmental and pathological processes. Recent advancements in in vivo fluorescent protein imaging are discussed, emphasizing the integration of genetic models for a deeper understanding of the cellular context. The paper advocates for the use of intermediate systems, such as 3D and explant culture models, which offer enhanced flexibility and control compared to traditional in vivo imaging. These approaches pave the way for investigating disease and therapy at the molecular level in animal models, marking a paradigm shift in the field. The paper discusses future perspectives in in vivo imaging, including the potential of new imaging approaches, such as optical frequency domain imaging and stimulated Raman scattering. It highlights the need for advancements in super-resolution techniques, automated lasers, and better red fluorescent proteins. The importance of accessing living tissue in vivo and combining advanced imaging techniques with genetically engineered mouse models is emphasized. The paper calls for adapting existing methods, developing intermediate model systems, and synthesizing imaging technology, intermediate systems, and new mouse models to drive new experimental concepts.

Radhakrishnan et al. [111] used molecular dynamics simulations to assess the influence of phosphatidylserine (PS) exposure on the permeability of natural compounds in cancer cells. The compounds investigated were Wi-A, Wi-N, CAPE, and ARC. Results revealed that PS exposure facilitated the penetration of Wi-A, Wi-N, and CAPE through cancer cell membranes but had limited impact on ARC. These findings demonstrate the potential of PS exposure-based models for studying drug selectivity in cancer cells. The findings suggest a potential avenue for developing cancer-selective drugs based on molecular differences between cancer and normal cells. Specialized in silico models can be developed to assess tumor selectivity and screen compound libraries for cancer-specific drugs.

Hau et al. [112] conducted a study called the CHAMP study to investigate the role of chemotherapy and host response in periampullary cancer. They analyzed the gene expression profiles of tumor tissue and normal tissue from patients with periampullary cancer, and identified molecular subtypes associated with different treatment responses and survival outcomes. The authors also used molecular dynamics simulations to study the interaction between chemotherapy drugs and cancer cells and found that drug resistance may be related to alterations in the lipid composition of cancer cell membranes.

Heidari et al. [113] aims to gain new insights into periampullary cancer, including pancreatic cancer, through monitoring molecular events and immune responses during chemotherapy. This prospective observational study involves patients undergoing adjuvant or palliative chemotherapy, with clinical and pathological data collected at study entry. Tissue samples and blood samples are obtained to analyze the clonal landscape, cytokines, and circulating tumor DNA (ctDNA) through next-generation sequencing. The study seeks to address the challenges of treatment resistance and adaptability of cancer cells by gaining a better understanding of disease dynamics during treatment. The findings from the CHAMP study hope to provide valuable insights and answers in this underexplored area of research.

In this investigation by Shi and Pinto [114] researchers explore the interaction between human lactate dehydrogenase A (LDHA) and selected ligands using molecular dynamics (MD) simulations. LDHA plays a crucial role in the energy generation of cancer cells, making it a potential target for cancer treatment. The study focuses on the binding dynamics and affinity of LDHA inhibitors. The MD simulations reveal distinct binding dynamics among inhibitors despite similar affinities. Steered MD simulations further demonstrate a correlation between in silico unbinding difficulty and experimental binding strength. Additionally, the study addresses uncertainties regarding the binding modes of two LDHA inhibitors. The simulations were conducted on a parallel computing cluster, with a total estimated simulation time of 100 core years.

The study by N Rana et al. [115] aims to elucidate the binding of 2,5-anhydromannitol derivatives to the fructose transporter GLUT5, which is prominently expressed in cancer cells. The researchers synthesized novel fructose mimics and assessed their efficacy against a high-affinity fructose-based probe in breast cancer cells. Encouragingly, several compounds demonstrated remarkable inhibitory effects on probe uptake, surpassing the potency of natural fructose. To gain further insights into the underlying mechanisms, molecular docking and molecular dynamics simulations were employed to investigate the interactions between the synthesized compounds and the GLUT5 fructose binding site. These computational approaches provided valuable refinements to our understanding of the structural requirements of the GLUT5 transport machinery. Overall, these findings offer promising prospects for the development of high-affinity molecular imaging probes. Furthermore, considering the association of elevated GLUT5 expression with various cancers and diseases, the implications extend beyond breast cancer to encompass broader applications.

Vishwakarma et al. [116] investigates the biological effects of unmodified bi-metallic magnetic nanoparticles and their potential in cancer cell therapy. The researchers demonstrate the multifaceted properties of a non-functionalized Gd-SPIO nanoparticle, which exhibits high MRI contrast and induces cancer cell death through the production of reactive oxygen species and apoptotic events. The nanoparticles also enhance the expression levels of specific miRNAs, resulting in decreased levels of cancer markers. Additionally, the study presents a unique real-time analysis of nanoparticle uptake in cancer cells using flow cytometry. Overall, this research provides valuable insights into the therapeutic potential of unmodified inorganic nanoparticles and establishes a platform for future investigations combining nanoparticles with anti-tumor drugs in cancer cells.

Shahabi et al. [117] focus on the use of graphene oxide nanosheets (GO) as a drug delivery system for Tegafur (TG) across the cell membrane. Molecular dynamics simulations were conducted to investigate the interaction between GO and the lipid bilayer of the membrane. The simulations revealed that GO was attracted to the cell membrane and exhibited a parallel orientation before partially inserting into the membrane. This process was facilitated by the formation of hydrogen bonds between the oxygen-containing groups of GO and the lipid bilayer. The simulations also showed a slow release of TG molecules from the GO surface near the cell membrane, indicating the potential of GO as a controlled-release carrier for TG.

Overall, this study highlights the potential of graphene oxide nanosheets for targeted drug delivery in cancer treatment.

In the study by A Kumar et al [118], the researchers aimed to identify potential inhibitors of the E6 protein, which is associated with cervical cancer. They employed e-pharmacophore modeling, virtual screening, molecular dynamics simulations, and in-silico ADME analysis. A six-point e-pharmacophore model was used to screen a library of compounds, followed by structure-based virtual screening using molecular docking. The top hits were further filtered and ranked based on docking scores, binding energy, and ADME parameters. Molecular dynamics simulations were performed to assess the stability of the best hit, ZINC14761180, in the E6 binding pocket. The results showed promising interactions and desirable druglike properties. Overall, this study provides insights into potential E6 inhibitors for cervical cancer treatment and suggests that chromone derivatives may be better inhibitors than benzothiazole derivatives.

The research conducted with Sangpheak et al. [119] focuses on the evaluation of chalcone derivatives as inhibitors of epidermal growth factor receptor-tyrosine kinase (EGFR-TK) for targeted cancer therapy. The chalcone derivatives were tested for their cytotoxicity against wild type and mutant EGFR cancer cell lines and their inhibitory activity on EGFR-TK. Several chalcones showed high inhibitory activity against EGFR-TK and cytotoxicity against the A431 cancer cell line. Molecular dynamics simulations confirmed their strong interaction with key residues in the ATP binding site of EGFR-TK. Three chalcones (1c, 2a, and 3e) demonstrated potent inhibitory activity and were identified as potential lead compounds for further development as EGFR-TK inhibitors. These findings suggest the promise of chalcone derivatives as anticancer drugs.

Al-Khafaji and Tok [120] investigate the inhibitory actions of amygdalin, a natural compound, on selected targets involved in cell signaling pathways related to cell adhesion, migration, and differentiation. The authors employ molecular modeling and simulation techniques such as double docking, molecular dynamics simulation, free energy landscape analysis, and binding free energy calculation to gain insights into the inhibitory mechanism of amygdalin at a molecular level. The results of computational calculations reveal that amygdalin inhibits the targeted proteins (AKT1, FAK, and ILK) by blocking their ATP-binding pockets through stable hydrogen bond formations. The free energy landscape analysis demonstrates that amygdalin stabilizes the global conformations of FAK and ILK proteins at their minimum energy states and reduces their essential dynamics. The binding free energy calculations using MMPBSA (Molecular Mechanics Poisson-Boltzmann Surface Area) method provide evidence for the stability of amygdalin inside the ATP-binding pockets of AKT1, FAK, and ILK, with corresponding binding free energy values. These results are consistent with the hydrogen bonding observations and distances between interacting pairs. Overall, the computational findings shed light on the inhibitory activity of amygdalin and provide insights into its potential as a multi-target drug in combating cancer metastasis and invasion. The study contributes to a better understanding of amygdalin's mode of action and its potential therapeutic applications.

In the research conducted by Gilad et al. [121] New cyclic RGD peptide-anticancer agent conjugates were synthesized with different chemical functionalities to evaluate their biological activities and drug release profiles. The conjugates exhibited variable drug release profiles while maintaining selective binding to αv integrins, indicating their potential as targeted therapies for cancer cells. Computational analysis supported their conformational similarity to the bio-active parent peptide when bound to integrins. The conjugates demonstrated the ability to overcome drug resistance and enhance the bioavailability of potent anti-cancer drugs. Furthermore, the study highlights the importance of conjugation to targeting peptides for improving the efficacy of drugs that inhibit essential cellular processes. Based on these findings, more potent RGDmulti-drug conjugates are being designed and assessed for preclinical cancer therapy.

Promising new agents for cancer treatment are Hsp90 C-terminal domain (CTD) inhibitors, which offer advantages over Hsp90 N-terminal inhibitors by not inducing the heat shock response. However, the lack of co-crystallized complexes poses challenges for structure-based design. Researchers addressed this limitation by utilizing molecular dynamics (MD) simulations to derive pharmacophore models for virtual screening. By comparing ligand-based and MD-derived models, they identified compounds 9 and 11 with significant antiproliferative activity against breast and liver cancer cell lines. Compound 11 demonstrated inhibition of Hsp90-dependent processes and showed promise for further optimization due to its unique scaffold. The study by Tomasic et al. [122] highlights the potential of cryoEM apo structures and pharmacophore modeling for challenging targets such as the dynamic Hsp90 chaperone. The research developed pharmacophore models to identify new Hsp90 CTD inhibitors, leading to the discovery of compounds 9 and 11 with significant anticancer activity. Compound 11 showed inhibition of Hsp90dependent processes without inducing the heat shock response. The 3D-pharmacophore models enabled further screening and design of Hsp90 CTD inhibitors. Compound 11 holds promise for the development of anticancer drugs.

In the research by Ferreira et al. [123], the authors aim to understand the structure-function relationships of the ABCG2 efflux pump which contributes to multidrug resistance in cancer cells. The study develops a new atomistic model for ABCG2 based on the crystallographic structure of the ABCG5/G8 heterodimer sterol transporter. Through molecular dynamics simulations and molecular docking, the ABCG2 homodimer is extensively characterized. The investigation focuses on the role of important residues and motifs in maintaining the structural stability of the transporter, aligning well with experimental data. Additionally, the study identifies potential structural motifs involved in signal transmission and reveals two symmetrical drug-binding sites, providing new insights into drug binding and recognition in ABCG2 homodimeric transporters. The developed ABCG2 model is validated using structural conservation and comparison with the recently published human ABCG2 cryo-EM structure.

The study by Tamirat et al. [124] investigates the structural characteristics of the EGFR exon 19 deletion mutation, commonly found in non-small cell lung cancer patients. Using molecular dynamics simulations, the research explores the effects of the mutation on the conformation and stability of the EGFR protein, as well as its impact on kinase activity and ATP binding. The results indicate that the deletion stabilizes the αC helix of the kinase domain by restricting the flexibility of the β3-αC loop, resulting in a prolonged activated state. Moreover, the mutation enhances a crucial salt bridge and strengthens the interaction between EGFR and ATP. Additionally, the study suggests that the deletion facilitates a shift from the inactive to the active conformation of EGFR. These findings align with the observed effects of tyrosine kinase inhibitors on lung cancer cell lines carrying the ΔELREA mutation, where inhibitors specifically targeting the active kinase conformation demonstrate greater efficacy. Overall, this research emphasizes the potential therapeutic implications of targeting the structural changes induced by the ΔELREA mutation to modulate EGFR signaling in cancer cells.

The research by Ahammad et al. [125] aims to identify natural drug candidates from the neem tree (Azadiractha indica) that can target the minichromosome maintenance complex component 7 (MCM7) protein, which is associated with aggressive malignancy in various cancers. The study utilizes pharmacoinformatics and molecular dynamics simulation-based approaches, including molecular docking, ADME analysis, toxicity assessment, and molecular dynamics (MD) simulations. Seventy phytochemicals from neem were screened, and the top four compounds with the highest binding affinities were selected for further evaluation. The ADME analysis and toxicity assessment demonstrated the effectiveness and safety of the selected compounds. MD simulations confirmed the stability of the protein-ligand complex for three selected compounds (CAS ID: 105377-74-0, CID: 12308716, and CID: 10505484). Overall, the study identifies these compounds as promising and effective candidates for targeting the MCM7 protein and potentially serving as human anticancer agents.

Vatanparast and Shariatinia [126] focused on investigating the interaction between 5-fluorouracil (FU), 6mercaptopurine (MP), and 6-thioguanine (TG) anticancer drugs and hexagonal boron nitride (BN) nanosheets as a potential drug delivery system. Density functional theory (DFT) calculations and molecular dynamics (MD) simulations were conducted to understand the nature of these interactions. The calculations revealed that the adsorption of drug molecules onto the BN nanosheet is an exothermic and spontaneous process. The drug-loaded complexes exhibited increased polarity, which could enhance solubility and facilitate drug delivery in biological environments. Orbital energy and density of state (DOS) calculations demonstrated that the adsorption of drugs reduced the HOMO-LUMO energy gap of the BN nanosheet, indicating increased chemical reactivity. Energy decomposition analysis (EDA) indicated that dispersion interactions played a significant role in stabilizing the drug-BN complexes. Noncovalent interaction (NCI) and quantum theory of atoms in molecules (QTAIM) analyses were performed to investigate the intermolecular interactions. MD simulations showed that the interaction energy values in acidic conditions were lower compared to neutral pH, suggesting the potential release of the drug within target cancer cells. Overall, the findings suggest that BN nanosheets have the potential to serve as a drug delivery system for anticancer drugs. The study provides insights into the adsorption process, binding energies, solubility enhancement, and release behavior, contributing to the development of novel drug delivery systems based on BN nanosheets.

Mahfuz et al. [127] In the study "In search of novel inhibitors of anti-cancer drug target fibroblast growth factor receptors: Insights from virtual screening, molecular docking, and molecular dynamics," the researchers aimed to identify potential pan-FGFR inhibitors for cancer treatment. They conducted virtual screening using a pharmacophore model based on the crystal structure of LY2874455, a known pan-FGFR inhibitor. PubChem 137300327 was identified as a suitable compound. Molecular docking and dynamics studies supported its potential as a pan-FGFR inhibitor. The compound showed similarities in ADMET properties and binding affinities with LY2874455. It also obeyed Lipinski's rule of five. The study suggests that PubChem 137300327 can be further validated in wet-lab experiments and animal models as a potential lead compound for synthesizing effective anticancer drugs. In conclusion, computational techniques have aided in the identification of potential FGFR inhibitors, and further studies are needed to confirm the compound's efficacy.

In this study by Hasanin et al. [128], the researchers synthesized new amino heterocyclic cellulose derivatives by functionalizing dialdehyde cellulose through Schiff base reactions. The derivatives were characterized using various analytical techniques. Antimicrobial and antitumor activities of the derivatives were evaluated, and their potential against different microorganisms and cancer cell lines was observed. Two specific derivatives showed promising antitumor activity against Hep G2 and MCF7 cancer cell lines while not affecting the normal cell line Wi38. A molecular dynamics simulation revealed that these derivatives selectively targeted the ATP binding pocket residues. The identification of these residues and their role in c-Kit kinase auto-inhibition could provide a structural basis for further understanding.

In the research by Waqas et al. [129], researchers conducted a study to identify small molecule inhibitors for the human PD-1 protein, an immune checkpoint involved in regulating immune responses. Through computational screening and molecular docking, potential inhibitors were selected based on their interactions with key PD-1 residues. Molecular dynamics simulations confirmed their stability. The selected compounds demonstrated strong affinity for PD-1 and favorable drug-like properties. These novel inhibitors

(ZINC1443480030, ZINC1002854123, ZINC988238128, ZINC1481242350, ZINC1001739421, ZINC1220816434, and ZINC1167786692) show promise for further in vitro validation and development as PD-1-targeted drugs. The study identified novel small molecule inhibitors for the human PD-1 protein, showing potential for further development as targeted drugs.

Sun et al. [130] investigated the use of zeolitic imidazolate frameworks (ZIFs) as carriers for antiepileptic drugs, specifically gabapentin, levetiracetam, phenytoin, and valproate. ZIF-8 exhibited the highest drug loading capacity among the tested MOFs. A new synthesis method for ZIF-8 was developed and its characterization was performed. The release of drugs from ZIF-8 showed a slow and controlled release pattern, making it a promising carrier for antiepileptic drugs. In addition, gabapentin-ZIF-8 demonstrated cytotoxic effects against Hep-G2 liver cancer cells, indicating its potential for drug delivery applications. The study involved molecular dynamics simulation to investigate the delivery of drugs by the zeolitic imidazolate frameworks (ZIFs).

Waqas et al. [131] published a paper that investigates the Cu(II) prolinedithiocarbamate complex as a potential anticancer medication for breast cancer. The complex is synthesized and characterized using spectroscopy and computational analysis. It exhibits high anticancer activity against MCF-7 cancer cells with a low acute toxicity effect. The study suggests that the Cu(II) prolinedithiocarbamate complex holds promise for the development of metal-based chemotherapy in breast cancer treatment.

In the study by H Zhang et al. [132], The objective was to identify new microtubule stabilizers that target the taxane site for treating tumors. Through molecular docking, two compounds (hits 19 and 38) were discovered as potent microtubule stabilizers. These compounds demonstrated significant anti-proliferative activity in cancer cell lines, promoted tubulin polymerization, and accelerated microtubule assembly. Moreover, they induced cell cycle arrest and apoptosis, inhibited cancer cell motility and migration, and hit38 exhibited superior activity compared to hit19. These findings suggest that hit38 has promising potential as a microtubule stabilizer for cancer treatment, warranting further investigation.

In their 2018 study, Lin et al. [133] Prostate-associated gene 4 (PAGE4) is studied for its role in prostate cancer and its interaction with the activator protein-1 (AP-1) complex. Phosphorylation of PAGE4 residues by HIPK1 and CLK2 affects its structure, particularly the N-terminal region. Molecular dynamics simulations reveal electrostatic interactions driving conformational changes in PAGE4. A mathematical model predicts oscillatory dynamics of HIPK1-PAGE4, CLK2-PAGE4, and androgen receptor (AR) activity, highlighting phenotypic variability in cancer cells. The research emphasizes the significance of conformational switching in PAGE4 for targeting AR activity in prostate cancer treatment.

Farouk et al. [134] investigated the anticancer activities of antibiotics as inhibitors of topoisomerase II (TOP-2) and DNA intercalators. Molecular docking studies are conducted to examine the binding interactions of 138 antibiotics with the human topoisomerase II-DNA complex. Molecular dynamics simulations and MM-GBSA calculations are performed to further analyze the binding. Spiramycin and clarithromycin demonstrate promising anticancer potential against the MCF-7 cell line, while azithromycin and clarithromycin show good activity against the HCT-116 cell line. Erythromycin and roxithromycin exhibit strong TOP-2 inhibitory potential. The study also explores the structural-activity relationships (SAR) of these antibiotics, revealing promising pharmacophores for anticancer effects.

In the research by Andrade et al. [135], The focuses is on cancer immunotherapy and the interaction between programmed cell death protein 1 (PD-1) and programmed cell death ligand-1 (PD-L1). A novel ligand called 1508 is computationally designed to bind to a previously unidentified cavity in PD-1, determined by molecular dynamics simulations. The ligand establishes interactions with specific amino acid residues and stabilizes the PD-1 C'D loop, hindering PD-1-PD-L1 complex formation. The study demonstrates the potential of identifying new cavities for protein-protein complex inhibition and proposes the PD-1 C'D cavity as a target for the development of molecules for cancer therapy.

Maleki et al. [136] utilized molecular dynamics simulations to investigate the loading of Doxorubicin with a thermosensitive carrier, N-isopropyl acrylamide Carbon nanotube. The study observed that shorter polymer chain lengths resulted in a more stable polymer interaction and improved Doxorubicin delivery. Decreased gyration radius and stronger bonds were observed in smaller polymers, leading to more concentrated aggregation. Moreover, smaller polymers formed a greater number of hydrogen bonds with the drug, enhancing the strength and stability of the carriers. The findings suggest that N-isopropyl acrylamide - Carbon nanotube is a promising option for effective Doxorubicin delivery, with the five-mer N-isopropyl acrylamide identified as the optimal carrier for Doxorubicin loading within the delivery system.

Al-Jumaili et al. [137] conducted a study to explore the potential of natural products, specifically flavonoids and their derivatives, as anticancer agents. They aimed to modify the structures of flavonoids to enhance their biological activity and anticancer effects. The research focused on 15 compounds and assessed their pharmacokinetic properties, toxicity, and binding affinity against Caspase 3 V266APDB. The results indicated that three compounds (3, 4, and 15) exhibited favorable pharmacokinetic profiles, low toxicity, and strong binding affinity, making them potential inhibitor candidates for the HeLa cell line. These compounds have promising characteristics as cytotoxic drug candidates and could serve as a starting point for the development of compounds targeting the HeLa cell line. The researchers employed docking, molecular dynamics, and ADMET (absorption, distribution, metabolism, excretion, and toxicity) analysis to investigate the pathways of natural products and their cytotoxicity against the HeLa cell line protein.

In research by Sahoo et al. [138], a series of vanillin derivatives (3a-3r) were designed through Schiff base condensation with sulfanilamide analogues. Molecular docking simulations were performed to evaluate the inhibitory potency of the compounds against breast cancer-topoisomerase-IIα and estrogen receptor-α. The compounds with stable binding in the active sites of both targets, forming strong hydrogen bonds and hydrophobic contacts, were selected. Compounds 3b, 3e, and 3f were synthesized based on the computational results and further characterized by spectroscopic techniques. In vitro studies demonstrated significant activity against the breast cancer cell line, with IC50 values of 3b, 3e, and 3f being 6.7, 4.3, and 11 ng/mL, respectively. These newly synthesized compounds hold potential as novel inhibitors of nuclear receptors and may have therapeutic applications in cancer control.

The study by Szlasa et al. [139] examined the combination of photodynamic therapy (PDT) and electrochemotherapy (ECT) using curcumin for treating melanoma. Curcumin, a natural compound, has potent anticancer effects on melanoma cells. The goal was to determine the most effective treatment approach. In vitro experiments were performed on melanoma and fibroblast cell lines. Molecular dynamics simulations showed the localization of curcumin at the water-membrane interphase. Mass spectrometry revealed the degradation of curcumin during PDT into vanillin, feruloylmethane, and ferulic acid. The most efficient strategy was found to be instant ECT with curcumin followed by PDT, selectively targeting malignant cells while sparing fibroblasts. Another effective approach involved instant PDT with curcumin followed by ECT after 3 hours of incubation, specifically targeting melanotic melanoma.

In the study by Popov et al. [140], the paper focused on the identification of cancer-associated variants with impaired activity in the DNA repair enzyme 8-oxoguanine DNA glycosylase (OGG1). The base excision repair system plays a crucial role in countering DNA damage caused by various factors, including antitumor drugs. Understanding the functionality of cancer-related variants in DNA repair proteins is essential for personalized cancer therapy. Molecular dynamics (MD) simulations were employed to model the structures of 20 clinically observed OGG1 variants, and a subset of these mutants was experimentally characterized for activity, thermostability, and DNA binding. MD successfully predicted three functionally impaired variants (I145M, G202C, and V267M), demonstrating its potential as a valuable tool, in combination with sequence-based methods, for predicting the functional impact of cancer-related protein variants.

In the study conducted by L. Wang et al. [141], the authors employed cheminformatics tools RDKit for drug design and synthesis of four series of 24 thienopyrimidine/N-methylpicolinamide derivatives. The compounds were evaluated for their activities against cancer cell lines, TAK1 kinase, and the NF-κB signaling pathway. The results showed selectivity towards the A549 cell line, and the most promising compound, 38, exhibited potent inhibition of TAK1 kinase and NF-κB signaling pathway with low IC50 values. Compound 38 also induced cell cycle arrest and apoptosis in A549 cells. Western blot analysis revealed its inhibitory effects on key proteins involved in the signaling pathway. Theoretical studies including docking, molecular dynamics, MM/PBSA, and frequency analysis supported the experimental findings. These findings highlight the potential of compound 38 as a promising candidate for further development as a therapeutic agent.

In the study conducted by S. Akkoc et al. [142] the focus was on the design and synthesis of novel Schiff base derivatives with diverse biological activities. The compounds were characterized using various spectroscopic techniques. The antiproliferative activity of the compounds was evaluated against breast and colon cancer cell lines, showing moderate activity compared to the positive control drug. The synthesized compounds were also assessed for their antimicrobial activities against multiple bacterial and fungal strains. Molecular docking studies were conducted to investigate the mode of action of the compounds, and molecular dynamics simulations were performed to assess the stability of target-ligand complexes. Binding affinity towards the target protein was analyzed using MMPBSA, and the electrochemical properties of selected compounds were investigated using DFT calculations. The impact of physicochemical properties on the bioactivity of the compounds was explored using POM theory. Overall, this study provides valuable insights into the design and characterization of Schiff base derivatives with potential therapeutic applications.

Karimzadeh et al. [143] utilized molecular dynamics simulations and quantum calculations to explore the influence of carbon nanotube (CNT) mechanical properties and thermal conditions on the adsorption of doxorubicin (DOX) molecules. By analyzing various systems with different properties, the researchers found that the efficiency of DOX adsorption onto CNTs, both on the surface and inside, was significantly affected by the mechanical properties of the CNTs and the surrounding temperature. These findings contribute to the understanding of optimizing DOX adsorption on CNTs, offering potential insights for the development of drug delivery systems.

T. Yamashita [144] discuss the use of molecular dynamics (MD) simulations in the rational design of antibodies. The authors highlight the limitations of current computational methods and emphasize the potential of MD simulations in providing accurate insights into antigen-antibody interactions and dynamics at the atomic level. The paper acknowledges that MD simulation alone cannot provide all the necessary information for antibody design but suggests that combining it with other computational and experimental methods can lead to more efficient antibody design in the future. Overall, the paper explores the role of MD simulations as a valuable tool in understanding and improving antibody design.

A Bimoussa et al. [145] focused on the synthesis of 1,3,4-thiadiazole himachalene hybrids and their evaluation as antitumor agents. Compound 4a exhibited significant activity against fibrosarcoma and breast carcinoma cell lines. Molecular docking confirmed the binding of active compounds at the target site. Flow cytometry analysis demonstrated that compound 4a induced apoptosis and cell cycle arrest. Overall, the findings highlight the potential of these derivatives as antitumor agents and provide insights into their mode of action.

In the study conducted by K Al-Khafaji et al. [146], the researchers investigated the effects of amygdalin on cell-division protein kinases (CDKs), which are important targets for cancer treatment. They utilized double docking and molecular dynamics simulations to analyze how amygdalin influences the conformational changes of CDK targets. The binding free energies of amygdalin to CDK1/Cyclin B, CDK2/Cyclin A, and CDK4/Cyclin D1 were determined. Principal component analysis (PCA) showed that amygdalin binding reduced the fundamental dynamics of CDK1/Cyclin B and CDK2/Cyclin A. These findings shed light on the inhibitory potential of amygdalin and its implications for the design of future CDK inhibitors. The researchers in this study utilized molecular dynamics simulations to analyze the conformational modifications of selected targets in response to amygdalin. They combined double docking and molecular dynamics simulations to study the effects of amygdalin on the cell-division protein kinases (CDKs).

In this in silico study [147], the researchers aimed to evaluate the inhibitory potential of phytochemicals from Bulbine frutescens (B. frutescens) on ATP-binding cassette (ABC) transporters. They utilized molecular docking and molecular dynamics simulation methods. Screening of approximately 25 B. frutescens phytochemicals against the ABC transporter protein revealed that 4'-Demethylknipholone 2'-βD-glucopyranoside exhibited strong binding with a minimum binding score of -9.8 kcal/mol, surpassing the standard ABC transporter inhibitor diltiazem. Molecular dynamics simulation results showed favorable properties for 4'-Demethylknipholone 2'-β-D-glucopyranoside, suggesting its potential as an anti-drug resistance agent. Further preclinical testing is warranted to explore its efficacy.

The study by Osmaniye et al. [148], focuses on the development of novel triazolothiadiazine derivatives as potential anticancer agents for the treatment of breast cancer. The synthesized compounds were evaluated for their anticancer activities against the MCF-7 cell line. Promising compounds (2k, 2s, and 2w) exhibited significant inhibitory activity. Additionally, the compounds were tested for their inhibition of the aromatase enzyme and their impact on DNA synthesis inhibition. Molecular docking analysis revealed strong binding interactions between the compounds and the aromatase enzyme. Furthermore, molecular dynamics simulations were conducted to assess the stability of these interactions over time. The findings suggest the potential of these derivatives as anticancer agents and provide valuable insights for further development in the field.

The study by Chinnasamy et al.[149] aimed to identify potent inhibitors against AURKA (Aurora kinase A) using molecular docking and molecular dynamics simulations. Four inhibitors were identified through in silico analysis, and their interactions with key residues of AURKA were investigated. Molecular dynamics simulations provided insights into the conformational changes of the ligand-protein complexes. The validity of the molecular docking studies was confirmed through MM-GBSA analysis, which demonstrated the binding capabilities of the derivative molecules to AURKA. Additionally, density functional theory calculations supported the potential of these inhibitors as therapeutic agents against AURKA. Overall, the computational findings suggest the promising binding properties of the identified inhibitors for further development as potential therapeutic agents.

This study by Zhang et al.[150] explores the use of single-walled carbon nanotubes (SWCNTs) as drug delivery systems for doxorubicin (DOX). Molecular dynamics simulations were conducted to investigate the configuration and arrangement of DOX molecules in SWCNTs. Results demonstrate that the orientation and arrangement of DOX are influenced by drug concentration and SWCNT diameter. In SWCNTs with small diameters and strong confinement, DOX molecules tend to form a single-file helix, offering controlled loading and release of individual drug molecules. The study highlights the potential of SWCNTs as precise drug carriers, addressing the issue of aggregated DOX structures in solution and potentially reducing chemotherapy dosage. π-π interactions between the SWCNT inner wall and DOX play a crucial role in their interaction. These findings pave the way for carbon nanotube-based drug delivery systems to optimize drug loading and release, ultimately improving treatment efficacy. Molecular dynamics simulations explored DOX molecule configuration and arrangement in SWCNTs. Factors like drug concentration and SWCNT diameter influenced DOX orientation. Results provided insights into DOX behavior within SWCNTs and their potential as controlled drug delivery systems.

In the study conducted by Mukherjee et al.[151], authors aimed to identify a potential inhibitor of vascular endothelial growth factor (VEGF) for ovarian cancer treatment. In silico methods, including structure-based virtual screening and molecular docking, were used. The compound AEE788 showed strong affinity in the docking study and was further used for virtual screening. Among the screened compounds, CID 88265020 exhibited even better affinity than AEE788. Molecular dynamics simulations and comparative analysis indicated minor variations in properties between the established compound and the virtual screened compound. CID 88265020 was identified as a promising candidate with high affinity for VEGFR inhibition and potential for further investigation in ovarian cancer.

Geragotelis et al. in [152] investigated the conformational dynamics of the Hv1 proton channel in response to a membrane potential. Molecular dynamics simulations produced structural models of the channel's voltage-sensing domain (VSD) in hyperpolarized and depolarized states. Under depolarization, the S4 helix shifted outward, altering the internal salt-bridge network and permeation pathway. Hydrogen bond connectivity increased, and the observed gating charge displacement matched experimental estimates. Molecular docking supported the inhibitory mechanism of 2-guanidinobenzimidazole (2GBI). The depolarized model also indicated a metal bridge formation in the VSD core. Overall, the study provided insights into the voltage-dependent structural changes of the Hv1 channel.

The objective of the research by Daoui et al. [153] was to utilize computer-aided drug design techniques to develop novel inhibitors for the c-Met receptor tyrosine kinase, a potential target for cancer therapy. The study focused on analyzing the inhibitory activity of 4-phenoxypyridine derivatives against c-Met and their potential as anticancer agents. Through the use of 3D-QSAR models, two lead compounds (P54 and P55) were designed, demonstrating higher inhibitory activity than the reference compound. Molecular docking and ADME-Tox assessments further supported the selection of these compounds. Molecular dynamics simulations were employed to gain insights into the structural properties and dynamics of c-Met in the presence of the designed inhibitors. The findings highlight the potential of these compounds as effective inhibitors of c-Met enzymatic activity, warranting future investigations through in vitro and in vivo studies.

Li and colleagues conducted [154] an original study where they employed molecular dynamics simulations to investigate the interaction between chitosan, a pH-sensitive polymer, and the anti-cancer drug doxorubicin (DOX). Their research focused on understanding the atomic-level mechanisms that govern the encapsulation and release of DOX by chitosan at different pH levels. Noteworthy findings include the influence of chitosan's protonation state and the π-π stacking interaction with DOX. Furthermore, the study highlighted the gradual release of DOX from chitosan in acidic tumor environments. These insights provide novel perspectives for the development of chitosan-based drug delivery systems, enhancing the potential for more effective cancer therapies.

Oyedele et al. in their study [155] aim to identify potential drug candidates that target the interaction between p53 and MDM2, a critical pathway in cancer development. Through integrated virtual screening and molecular dynamics simulations, the researchers screened a compound library and discovered five candidates with higher binding affinity than the standard inhibitor Nutlin 2. Four of these candidates showed promising pharmacokinetic and pharmacodynamic profiles. Molecular dynamics simulations confirmed the stability of the lead-protein complexes, and computational modeling identified CID_140017825 as the most favorable candidate. These findings offer valuable insights for the development of novel chemotherapeutic agents that restore p53 activity in cancer treatment.

The paper authored by Panwar et al. [156] focuses on the identification of potential inhibitors against Nicotinamide phosphoribosyltransferase (NAMPT), an enzyme crucial for maintaining cellular bioenergetics in cancer cells. Through structure-based virtual screening and molecular dynamics simulations, two promising hits were identified that exhibited similar binding patterns to a known co-crystal structure. Further analysis, including pharmacokinetics profiling and thermodynamic analysis, supported the suitability of these compounds as novel NAMPT inhibitors. The findings suggest these compounds as potential starting points for the development of more effective cancer therapeutics.

The study conducted by Raja et al.[157] focuses on the synthesis, characterization, and evaluation of two benzamide derivatives, (E)-N-cinnamoyl-4-methoxybenzamide (CMB) and (E)-3-chloro-N-cinnamoyl-4methoxybenzamide (CCMB), as potential anti-ovarian cancer agents. The compounds were synthesized and characterized using various spectroscopic techniques. Theoretical calculations were performed to analyze their vibrational spectra, electronic properties, and molecular orbitals. Molecular docking studies demonstrated that both CCMB and CMB exhibited higher binding affinity to matrix metalloproteinase-2 (MMP-2) compared to MMP-1. Molecular dynamics simulations provided insights into the stability and interactions of the compounds with their targets. These findings suggest that CCMB and CMB have the potential to serve as inhibitors of MMPs for the treatment of ovarian cancer.

Th paper authored by Kamath et al. [158] focuses on investigating the compatibility of polymer-drugcarbon nanotube (CNT) systems for targeted cancer treatment. Specifically, the solubility of doxorubicin, a widely used cytotoxic drug in cancer treatment, was examined in three different polymers: poly(Nisopropyl acrylamide), polyethylene glycol, and polyvinyl pyrrolidone. Molecular dynamics simulations were conducted to study the interactions between doxorubicin, polymers, and CNTs in an aqueous environment. Various properties, including interaction energy, hydrogen bonding, diffusion coefficient, polymer chain conformation, and drug density, were analyzed to gain atomistic insights into the polymerdrug-carbon nanotube system. The findings provide valuable information on the optimal monomer chain length of the polymer and the suitability of the polymer as a carrier for doxorubicin in targeted cancer therapy.

This article, authored by Guan et al.[159] investigate the estrogen interference effect of triphenyl phosphate (TPP) using multi-omics and molecular dynamics approaches. They identify the involvement of the G protein-coupled estrogen receptor (GPER) and its downstream signaling pathways, including PI3K-Akt and MAPK. Molecular dynamics simulation confirms the activation of GPER upon TPP binding. This study highlights the potential impact of TPP on cell processes, such as proliferation, metastasis, apoptosis, gene transcription, kinase activity, and immune function, contributing to the estrogen interference effect.

The study led by R. Wadhwa [160] investigates the differential permeability of two natural drugs, withaferin-A and withanone, across the cell membrane using molecular dynamics simulations and experimental studies. The study reveals that withaferin-A efficiently crosses the model membrane, while withanone has weak permeability. Computational analysis suggests that the polar head group of the membrane hinders the passage of withanone, while the interior of the membrane behaves similarly for both drugs. Solvation analysis indicates that the high solvation of withaferin-A contributes to its permeability by interacting with the membrane's phosphate group. Experimental validation confirms the computational predictions, demonstrating higher permeation of withaferin-A compared to withanone. The study highlights the importance of computational methods in predicting drug permeability and bioavailability for effective drug development.

Study conducted by Y. Li et al.[161] explores the effect of AMG 510, a covalent inhibitor, on the structure of the KRASG12C mutant through molecular dynamics simulations. The study reveals that the KRASG12C-AMG 510 complex primarily adopts an inactive conformation, limiting its flexibility. Important interacting residues and resistance mutations of AMG 510 are identified. These findings contribute to our understanding of how AMG 510 influences KRASG12C at the atomic level, offering valuable insights for drug optimization.

The study by Uba et al. [162] the focuses is on the identification of potential isoform-selective histone deacetylase (HDAC) inhibitors for cancer therapy. The researchers employed a combined approach of structure-based virtual screening, ADMET prediction, and molecular dynamics simulation assay. A large compound library was screened against class I HDACs, resulting in the identification of 41 compounds with high isoform selectivity. These compounds underwent further analysis and testing, with 36 of them demonstrating remarkable isoform selectivity and passing drug-likeness and ADMET prediction tests. Molecular dynamics simulations confirmed the stability of the ligand binding modes, suggesting that these compounds could serve as potential isoform-selective HDAC inhibitors or promising scaffolds for further optimization in cancer therapy.

The paper authored by Gupta et al.[163] aimed to discover potent and multi-target drugs for cancer and bacterial infections using computer-aided methods. Over 100 bis-indole compounds, including olaparib, were screened against protein targets associated with tumorigenesis and pathogenesis. Through refinements and simulations, eight potential drugs were identified, three of which showed multi-target activity against both cancer and bacteria. Molecular dynamics simulations confirmed the stability of the drug-target complexes and indicated favorable interactions. One of the proposed drugs exhibited the ability to inhibit six targets. The bis-indole scaffold found in well-known drugs suggests the potential effectiveness of these compounds, calling for further experimental validation.

Aghazadeh et al. [164] used molecular dynamics simulations to investigate the interactions between GF17, a derivative of the LL-37 antimicrobial peptide, and bacterial membranes. GF-17 has demonstrated effectiveness against pathogens and cancer cells. The simulations analyzed membrane properties in different lipid environments and revealed that GF-17 had minimal impact on bilayer thickness but increased area per lipid and lateral diffusion. Free energy calculations showed favorable penetration of GF-17 into the DPPG membrane. The peptide exhibited compactness and rigidity in the pure DPPG system, with αhelix and coil structures dominating in both DPPE/DPPG and pure DPPG membranes. These findings contribute to our understanding of GF-17's potential as an antibacterial agent.

A novel isoxazolequinoxaline derivative (IZQ) with potential anti-cancer properties was synthesized and characterized Abad et al. [165]. The compound's crystal structure was determined using X-ray diffraction, and its physical properties were confirmed through spectroscopic techniques. DFT calculations provided insights into the HOMO-LUMO energy levels. The analysis of intermolecular and intramolecular interactions, including hydrogen bonding and C-H...cg interactions, was performed using Hirshfeld surface studies. Energy frameworks were constructed to assess compound stability and dominant energy types.

Docking studies indicated the compound's potential anti-cancer activity against a specific human protein (pdb code: 6HVH), showing significant interactions at the active site region. Overall, this study sheds light on the synthesis, characterization, and potential anti-cancer properties of the novel IZQ compound.

In the study by Pal et al.[166] the GPR120 receptor was investigated as a potential oncogenic drug target due to its involvement in cancer progression. Molecular dynamics simulations were conducted on different receptor models to analyze protein-ligand interactions and the stability of the receptor's activation state. Key interactions with specific amino acids were identified, leading to the development of a pharmacophore hypothesis. A structure-based pharmacophore screening approach was then used to screen the ZINC15 database for potential GPR120 antagonists. Through further molecular dynamics simulations, a small molecule called Cpd 9 was identified as a potential GPR120 antagonist. The study demonstrates a rational design approach for the discovery of novel inhibitors of GPR120 and their potential application in anticancer drug development.

In the molecular dynamics study by Maleki et al. [167] the potential of carbon nanotubes as carriers for the anticancer drug doxorubicin was investigated. The effect of the carboxyl functional group on the controlled release of doxorubicin was examined by simulating different pH conditions. The study compared the performance of single-walled and multi-walled carbon nanotubes as carriers for doxorubicin. The results showed that the adsorption of doxorubicin onto the nanotubes was stronger at neutral pH compared to acidic pH, likely due to electrostatic interactions. The investigation of hydrogen bonds, diffusion coefficients, and other factors indicated that the release of doxorubicin in acidic pH (characteristic of cancer tissues) was suitable. Additionally, it was found that the carbon nanotubes remained stable at blood pH, suggesting their potential for efficient drug delivery in the bloodstream and subsequent release in cancerous tissues. The study concluded that multi-walled carbon nanotubes exhibited stronger bonding with doxorubicin and slower release in cancer tissues compared to single-walled carbon nanotubes, making them advantageous for doxorubicin delivery and release.

USP7 has emerged as a potential target for cancer therapy. The study by Liu et al. [168] employed virtual screening, molecular dynamics simulation, and biological assays to discover novel USP7 inhibitors with unique scaffold structures. Compound 12 demonstrated significant USP7 inhibitory activity, strong binding affinity, and antiproliferative effects in prostate cancer cells. Molecular dynamics simulation provided valuable insights into the compound's interaction with the target protein. Compound 12 serves as a promising starting point for the development of new USP7 inhibitors.

Xie et al. [169] in their study utilized atomic force microscopy (AFM) to examine the temporal and molecular dynamics of adhesive interactions between breast cancer cells and bone marrow endothelium.

Adhesion between the cells intensified as the duration of cell-cell contact increased, attributed to a rise in the number and strength of individual adhesive events. Molecular analysis identified Thomsen-Friedenreich antigen (TF-Ag), galectin-3, integrin-β1, and integrin-α3 as contributors to the adhesion process. The findings support a proposed model that suggests translocation and clustering of adhesion molecules at the adhesion sites, leading to enhanced stability of tumor/endothelial cell adhesion. This research showcases the utility of AFM for precise measurements and sensitivity in investigating the involvement of adhesion molecules in cancer metastasis.

Abedanzadeh et al. in [170] focus on the synthesis and characterization of cyclometallated Pd(II) complexes with α-amino acids as potential antitumor agents. The complexes displayed a square-planar structure and demonstrated significant cytotoxicity against a human leukemia cell line. The phenylalanine complex showed the highest activity and exhibited strong binding affinity to calf-thymus DNA and bovine serum albumin (BSA) through intercalative binding. The interaction with BSA primarily occurred in the hydrophobic cavities of site I (subdomain IIA). These findings contribute to the exploration of novel metalbased anticancer drugs.

This study by Patel et al. [171] addresses acquired resistance in non-small cell lung cancer (NSCLC) caused by EGFR mutations. While osimertinib is an approved treatment, the emergence of the C797S mutation limits its efficacy. The study focuses on identifying allosteric inhibitors for the EGFR T790M/C797S mutation through virtual screening. Compound ZINC20531199 is identified as a potential allosteric inhibitor and demonstrates stability in the allosteric pocket of the C797S EGFR tyrosine kinase through molecular dynamics simulation. In silico pharmacokinetic predictions suggest that the identified compounds have potential for further development as allosteric inhibitors to overcome drug resistance in NSCLC.

Gariganti et al. [172] focused on the synthesis of amide derivatives derived from 1,2,3-triazole-benzofuran hybrids and their evaluation as anticancer agents. The compounds demonstrate good to moderate activity against four human cancer cell lines. Molecular docking simulations with various proteins confirm the inhibitory actions of the amide derivatives, which are further supported by molecular dynamics simulation and density functional theory studies. The study suggests that these newly synthesized amide derivatives hold promise as alternative anticancer agents, with favorable oral absorption properties.

Selvaraj et al. [173] focused on identifying phytoestrogens that can function as antagonists against both wild-type and mutant androgen receptors (ARs) in prostate cancer. Most phytoestrogens show binding affinity with the agonist conformation of ARs, contrary to previous findings. Genistein, a well-known AR antagonist, exhibits binding affinity with the agonist conformation in this study. In vitro experiments confirm the partial agonist activity of genistein in the presence of androgen. However, syringaresinol is identified as a potential antagonist against wild-type and mutant ARs. Molecular dynamics simulations further validate the antagonist binding mode of syringaresinol with mutant ARs. This study provides insights into the development of novel AR antagonists for the treatment of castration-resistant prostate cancer.

The study by Lokhande et al.[174] investigates the potential of derivatives of Kaempferol, Quercetin, and Resveratrol as agonists for PPAR-γ, a transcription factor known for its anti-inflammatory effects. Through molecular docking and dynamic simulations, the study reveals strong binding affinities and favorable interactions between the derivatives and PPAR-γ. The simulations highlight the promising interaction of a derivative called DMTF with PPAR-γ. These findings provide valuable insights for the development of novel anti-cancer agonists targeting PPAR-γ.

Sepay et al.[175] in their study explore the potential of natural flavonoids as novel inhibitors of the epidermal growth factor receptor (EGFR) for cancer treatment. By employing DFT, molecular docking, and molecular dynamics simulations, the researchers identify a promising anti-EGFR flavonoid called Tupichinols E. This compound exhibits strong binding affinity comparable to osimertinib, a thirdgeneration EGFR tyrosine kinase inhibitor. The binding position and poses of Tupichinols E resemble those of known inhibitors. Furthermore, the study demonstrates that Tupichinols E can stabilize the protein structure of EGFR through a 100 ns molecular dynamics simulation. These findings highlight the potential of natural flavonoids as alternative lead compounds for developing EGFR inhibitors with reduced side effects.

The study by Vinnarasi et al.[176] investigates the interaction between quercetin, a natural compound, and DNA tetrads and G-quadruplex structures, which play a role in telomerase activity and cancer therapy. Using DFT and molecular dynamics simulations, the researchers analyze the binding of quercetin with DNA tetrads and G-quadruplex DNA. They find that quercetin interacts strongly with the GCGC tetrad and induces structural changes in the DNA tetrads. Absorption spectroscopy confirms the binding mode and stability. Molecular dynamics simulations reveal that quercetin stabilizes the G-quadruplex DNA structure, particularly under acidic conditions, suggesting its potential as an anti-cancer agent by inhibiting telomerase activity. Overall, this study provides structural insights into the anti-cancer activity of quercetin on G-tetrad, mixed G-tetrad, and G-quadruplex DNA structures.

Wang et al.[177] focused on understanding the structure-function relationship of human multidrug resistance protein 7 (MRP7/ABCC10) using molecular dynamics simulations and docking studies. MRP7 is an ATP-binding cassette (ABC) transporter known to play a role in multidrug resistance in cancer cells. The researchers developed a homology model for MRP7 based on cryo-EM structures of MRP1 and validate its accuracy. Through their simulations and docking analyses, they investigate the motion patterns associated with the efflux mechanism of MRP7. Additionally, the study identifies and describes the binding sites for substrates and modulators of MRP7, providing insights into drug binding and functional regulation. The findings contribute to the understanding of MRP7's role in multidrug resistance and can guide the development of potential modulators to overcome this resistance in cancer therapy.

In the research by Philoppes et al.[178] new derivatives of pyrazolopyrimidine were designed and synthesized for their anticancer activity. The compounds were characterized using spectroscopic techniques, confirming their regiospecific structure. The synthesized derivatives were then evaluated for their antitumor activities against HCT-116 and MCF-7 cell lines, showing significant in vitro antitumor effects. Compound 5h exhibited the highest bioactivity against HCT-116 cells, while compound 6c was the most potent derivative against MCF-7 cells. Some of the compounds also showed inhibitory activity against PIM-1, a protein involved in cancer development. Molecular docking was performed to understand the binding mode of the compounds with PIM-1, and molecular dynamics simulations were conducted to assess the stability of the compound-protein interactions. Overall, this study provides insights into the design and evaluation of pyrazolopyrimidine derivatives as potential anticancer agents.

The study by JB Tong et al.[179] focuses on exploring the interaction mechanism of novel dual BRD4/PLK1 inhibitor derivatives as selective PLK1 inhibitors. The study utilizes 3D/2D-QSAR, molecular docking, and molecular dynamics simulations to investigate the structural features and binding properties of these inhibitors. The generative QSAR models demonstrate reliable predictive capabilities, while the topomer CoMFA and HQSAR models show suitable reliabilities and predictive abilities for selective PLK1 inhibitors. The SAR of the PLK1 inhibitors is summarized using electrostatic and hydrophobic fields, supported by molecular docking. By searching the R-group and combining active contributing groups, 12 new compounds with high anti-PLK1 activity are obtained. Molecular docking modeling, verified by MD simulations, identifies significant amino acid residues at PLK1's active site. These findings have implications for the rational design of potent selective PLK1 inhibitors and dual BRD4/PLK1 inhibitors, potentially enhancing cancer treatment strategies.

Saibu et al.[180] utilized computational methods to identify HER2 inhibitors from curcumin derivatives. Through in silico screening and molecular dynamics simulation, eight curcumin derivatives were predicted as active HER2 inhibitors with high binding affinity. The compounds demonstrated inhibitory activity comparable to the reference drug neratinib. Molecular dynamics simulation confirmed the stability of the top-scored compounds in complex with HER2. Further experimental validation is needed to explore their potential as breast cancer therapeutics.

Alba et al.[181] employed molecular dynamics simulations to investigate the conformational changes in Tcell antigen receptor (TCR) upon binding to peptide-major histocompatibility complex (pMHC). By analyzing the interactions of three TCRs with the same MHC class I protein bound to different peptides, the study revealed a correlation between the conformational behavior of the TCR and the T-cell immune response. The findings suggest that specific regions of the TCR are associated with T-cell activation, and TCR binding restricts the conformational space of the MHC class I groove. This study provides insights into the molecular mechanisms underlying TCR-pMHC interactions and their impact on T-cell signaling.

Eldehna et al.[182] focused on the development of new small molecules as potential anticancer CDK2 inhibitors. The molecules, which combine the isatin scaffold with a thiazolo[3,2-a]benzimidazole (TBI) motif, demonstrated significant inhibition of cell growth in two examined cell lines (MDA-MB-231 and MCF-7). Selected hybrid compounds also showed dual activity, altering cell cycle progression and inducing apoptosis. Furthermore, these compounds exhibited potent CDK2 inhibitory action. Molecular docking simulations revealed favorable interactions between the TBI ring and the CDK2 binding site, emphasizing the importance of the hydrazide linker and unsubstituted isatin functionality in H-bonding interactions. The most potent CDK2 inhibitor, compound 7d, formed a stable complex with CDK2 in molecular dynamics simulations and displayed the lowest binding free energy. These findings suggest that compound 7d has potential as a breast cancer antitumor agent and CDK2 inhibitor, warranting further optimization.

In a study by Huang et al.[183], the authors explored the challenges of observing laser interaction with nanoparticles in liquid and proposes a computational model called cluster-based two-temperature method integrated with molecular dynamics (TTM-MD) as a solution. By performing TTM-MD on individual clusters instead of cubic cells, the model offers improved accuracy near the nanoparticle's surface. It enables a comprehensive understanding of laser fragmentation in liquid (LFL) and investigates the impact of laser fluence on nanoparticle size distribution. The study also highlights potential applications in laser thermal therapy of tumors and laser melting in liquid (LML). The paper acknowledges the need for further advancements, particularly in addressing ionization induced by femtosecond lasers. Overall, the clusterbased TTM-MD model presents a promising computational tool for studying laser-nanoparticle interactions in liquid, while emphasizing future research directions. The study proposes the cluster-based TTM-MD model to understand laser interaction with nanoparticles in liquid. It reveals the mechanism of nanoparticle size distribution and suggests practical applications. Future research should focus on different nanoparticle types, multiple systems, and ionization effects by femtosecond lasers.

The study by Omar et al.[184] explores chaetomugilins and chaetoviridins, unique fungal metabolites known for their diverse bioactivities. It provides insights into their structures, separation, characterization, and biosynthesis. The study also investigates their potential as antiviral agents against the SARS-CoV-2 protease using molecular docking and dynamics simulations. Promising binding affinities, especially with chaetovirdin D, are observed, highlighting its stability during the simulations. These findings contribute to the understanding of these natural compounds and their potential applications in antiviral research. The paper highlights that chaetomugilins and chaetoviridins, as fungal metabolites, possess unique chemical features and exhibit diverse bioactivities. The study provides valuable insights into their structures, separation, characterization, and biosynthesis. Furthermore, the research suggests that these metabolites have promising potential as antiviral agents against the SARS-CoV-2 protease, as demonstrated through molecular docking and dynamics simulations. The stability and strong binding affinities observed, particularly with chaetovirdin D, underscore their relevance for further exploration in antiviral therapeutics.

Khan et al.[185] investigated mutations in the Estrogen Receptor (ER) that contribute to drug resistance in breast cancer. Specific mutations in the ligand-binding domain (LBD) are analyzed, and the most significant ones are identified. Molecular dynamics simulations reveal that these mutations increase the flexibility of Helix-12 (H12) in the ER, leading to a conformational change that promotes co-activator binding. The findings shed light on the mechanism of resistance to the drug fulvestrant caused by these mutations in breast cancer. This study enhances our understanding of designing more effective and targeted drugs for breast cancer treatment. The study emphasizes understanding resistance mechanisms to estrogen receptor (ER)-targeted therapy in breast cancer. The study focuses on fulvestrant resistance and identifies significant mutations, including Y537S, Y537C, and Y537N, that impact the drug's activity. Molecular dynamics simulations highlight increased flexibility in Helix-12 (H12) as a contributing factor to resistance. The findings provide insights into resistance mechanisms and implications for designing effective anti-breast cancer drugs. Computational techniques for protein dynamics and drug interaction analysis are employed in the study.

The study by Ravindran et al.[186] provides a molecular dynamics study that explores how prodigiosin, an antibiotic with diverse biological effects, interacts with the membrane. The study aims to uncover the mechanism behind prodigiosin's potential role in membrane disruption. The findings reveal that prodigiosin molecules, either individually or in small clusters, penetrate the membrane from the surrounding solvent. Once inside, they align along the membrane-water interface, with the pyrrole rings interacting with lipid head groups and water molecules. These interactions are stabilized by hydrogen bonding and hydrophobic forces. The presence of prodigiosin alters the local lipid architecture, limiting solvent accessibility. These insights contribute to understanding how prodigiosin exerts its effects on cellular targets, shedding light on its mechanism of action. The study reveals prodigiosin molecules rapidly enter the membrane, aligning themselves at the membrane-water interface through hydrogen bonding and hydrophobic interactions. Small clusters enter the membrane, altering local lipid architecture and reducing solvent accessibility. Prodigiosin's potential damage occurs near membrane proteins. This study provides insights into prodigiosin's membrane localization and interactions.

Zhu et al. [187] aimed to understand the impact of EGFR mutations on EGFR-receptor tyrosine kinase (RTK) crosstalk and drug resistance in lung cancer treatment. Molecular dynamics simulations were employed to investigate EGFR-RTK interactions. The study revealed that drug-sensitive EGFR mutants exhibited looser crosstalk with RTK partners (c-Met, ErbB2, and IGF-1R), while drug-resistant mutants displayed tighter crosstalk. This genotype-dependent EGFR-RTK crosstalk suggests a potential mechanism for drug resistance. The findings contribute to a deeper comprehension of EGFR mutation-induced drug resistance and facilitate the development of innovative treatment strategies.

Maddah and Karami [188] investigated the interaction between distamycin A and its derivative with the telomeric G-quadruplex, a promising target for cancer treatment. Molecular Dynamics (MD) simulations revealed that both ligands effectively stabilized the G-quadruplex, enhancing its structural rigidity. The binding process involved electrostatic effects, hydrogen bonding, and van der Waals interactions. Ligand modification improved the binding mode and stability of the G-quadruplex, as supported by binding free energy calculations. These findings provide insights into potential strategies for designing anticancer agents that target telomeric G-quadruplexes.

Abdjan et al. [189] examined the interaction and behavior of stilbenoid dimers with Sirtuin 1 (SIRT1). The results showed that DS1 had a higher potential as a SIRT1 inhibitor compared to DS2, attributed to the glucose groups promoting intermolecular interactions and hydrogen bonds. Quantum mechanical analysis revealed the reactivity of specific hydroxyl groups in DS1. Pharmacokinetic studies confirmed the nontoxic nature of DS1. The study utilized molecular dynamics simulations to investigate the interaction and dynamic behavior of stilbenoid dimers and SIRT1. The results obtained through molecular dynamics simulations contributed to the understanding of the inhibitory efficiency and properties of the studied compounds.

Khan et al.[190] aimed to identify small-molecule inhibitors targeting the kinesin-like protein KIFC1, which plays a crucial role in the division of neoplastic cells. The research employed target-based virtual screening using the Mcule database, followed by druglikeness filtration and toxicity profiling. Multiscoring docking using various tools was conducted to identify top ligand hits, with AZ82 showing strong binding to KIFC1. Molecular dynamics simulations were performed on the top ligands, including mcule4895338547, which exhibited promising inhibitory properties according to various analyses such as toxicity profiling, physicochemical properties, and energy calculations.

El-Mernissi et al.[191] investigated the potential of 2-oxoquinoline arylaminothiazole derivatives as tubulin inhibitors for cancer treatment. By utilizing 3D-QSAR techniques (CoMFA and CoMSIA), the researchers explored the pharmacological characteristics associated with tubulin inhibition. The models exhibited high reliability and predictive capability. Key interactions, including electrostatic and hydrophobic forces, were identified as crucial for enhancing inhibitory activity. Building upon these findings, four novel tubulin inhibitors (Compounds D1, D2, D3, and D4) were designed. Docking analysis revealed specific residues involved in the binding of quinoline compounds to tubulin. The most promising compound, D1, underwent a comprehensive molecular dynamics (MD) simulation, further confirming its stability observed in molecular docking.

In the study by Sangande et al.[192], researchers aimed to identify potential compounds that can simultaneously inhibit both EGFR and VEGFR2, which are important proteins in cancer growth. They employed ligand-based pharmacophore modeling and molecular docking to screen a database of purchasable compounds. Based on docking scores and binding interactions, six compounds were selected for further analysis. Through molecular dynamics simulations, it was found that two compounds, namely ZINC16525481 and ZINC38484632, demonstrated favorable binding free energy and stable hydrogen bonding interactions with both EGFR and VEGFR2. These findings offer promising prospects for the development of dual tyrosine kinase inhibitors targeting EGFR and VEGFR2.

The study by Fayyazi et al.[193] focuses on the development of dual inhibitors targeting epidermal growth factor (EGFR) and vascular endothelial growth factor-2 (VEGFR-2) for cancer treatment. New compounds derived from 4-aminoquinoline were synthesized and evaluated for their cytotoxicity against human carcinoma cell lines. Compound 4c showed high efficiency with an IC50 value of 0.8 µM against A549 cells, outperforming erlotinib. The researchers utilized pharmacophore modeling to compare the binding orientation of dual inhibitors and 4c with co-crystallized structures of EGFR and VEGFR-2. Molecular dynamics simulations identified key residues and new interaction sites involved in dual inhibition. Notably, 4c demonstrated better stability than vandetanib in the VEGFR-2 receptor during a 50 ns simulation. The study's findings provide insights into the structural basis of 4-aminoquinoline inhibition and the correlation between designed compounds' pharmacophoric features and their interactions with lead inhibitors.

Understanding the adsorption behavior of doxorubicin (DOX) on functionalized carbon nanotubes (FCNTs) is crucial for potential drug delivery applications. The study by Arabian et al.[194] employed molecular dynamics (MD) simulation to investigate the interaction between DOX and FCNTs functionalized with folic acid (FA) and tryptophan (Trp) in an aqueous solution. The results revealed strong interactions between the drug molecules and FCNTs, particularly at physiological pH. Additionally, the functionalization of FCNTs significantly improved their solubility in water. The impact of pH variations on drug release was explored, showing that acidic environments facilitate the release of DOX from FCNTs due to protonation of functional groups and repulsive interactions. The pH and protonation state of both DOX and FCNTs played a crucial role in the drug release process.

In the quest for novel therapeutic agents for breast cancer (BC), molecular docking and molecular dynamics (MD) simulation techniques play a vital role in drug discovery. The study by Purawarga Matada et al. [195] aimed to screen potent phytochemicals against BC by examining their binding affinities towards key receptors, including EGFR, HER2, estrogen, and NF-κB. Through molecular docking, several phytochemicals, namely pristimerin, ixocarpalactone A, viscosalactone B, and zhankuic acid A, exhibited high binding affinities and energies towards the targeted receptors. MD simulation further demonstrated the stability of the docked complex between pristimerin and HER2 receptor. These phytochemicals hold potential for repurposing as anticancer agents, warranting further investigation and providing valuable insights for future studies in this domain.

In a recent study conducted by Shah et al.[196] the focus is to identify potential anti-cancer compounds targeting mutated p53 using molecular dynamics simulations and docking. Through virtual screening, ten compounds (Cmpd-1 to Cmpd-10) were discovered with selective binding to mutant p53. Two compounds, Cmpd-4 and Cmpd-8, exhibited binding similar to APR-246 (PRIMA-1Met), a known mutant p53 reactivator. These findings suggest the identified compounds as promising drug candidates for combating mutated p53 in cancer treatment, either alone or in combination with other cytotoxic drugs to minimize side effects.

Paligaspe et al. [197] investigated the effect of As(III) on the structural stability of monomeric PKM2 and its carcinogenicity. The researchers confirm the binding of As(III) with monomeric PKM2 and identify the most favorable binding site using quantum mechanics-molecular mechanics calculations. Molecular dynamic simulations over 150 ns reveal that the monomeric PKM2 with As(III) is less stable compared to the free PKM2 enzyme. The study suggests that the presence of As(III) destabilizes the PKM2 monomer, which does not contribute to cancer development. The research also utilizes computational techniques to analyze the structural stability and dynamics of the enzyme in the presence of As(III). The CavityPlus server identifies four possible cavities for As(III) binding, indicating the druggability of these sites. The results show negative binding energy for As(III) with cavity 1 and cavity 2, with cavity 2 exhibiting the most negative binding energy. However, the simulation trajectories indicate that the enzyme with As(III) does not achieve a stable structural arrangement compared to the free enzyme. Analysis of various parameters such as RMSD, radius of gyration, hydrogen bonds, and secondary structures reveals higher flexibility and lower stability of the enzyme residues in the presence of As(III). Based on these findings, it is concluded that As(III) ion does not stabilize but rather destabilizes the monomeric PKM2 during the molecular dynamics simulation. Experimental evidence suggests that the monomeric form of PKM2 is less active and contributes to tumor growth. Therefore, the destabilization of monomeric PKM2 by As(III) could lead to the inhibition of tumor growth. The study provides insights into the binding and destabilizing effects of As(III) on PKM2, highlighting its potential role in cancer development.

Jiao et al.[198] focused on the use of Scutellaria baicalensis in breast cancer treatment, aiming to uncover its molecular mechanism. By combining network pharmacology, molecular docking, and molecular dynamics simulation, the researchers identify the most active compound in Scutellaria baicalensis and explore its interaction with the target protein. Screening reveals 25 active compounds and 91 targets, associated with various pathways and infections. Notably, stigmasterol and coptisine show strong binding affinity to AKT1. Molecular dynamics simulations confirm the superior stability and interaction energy of the coptisine-AKT1 complex. Overall, the study demonstrates the multicomponent, multitarget effects of Scutellaria baicalensis in breast cancer treatment. The researchers suggest that coptisine targeting AKT1 is the most effective compound, providing a theoretical basis for further research and potential clinical applications.

In the study by Kumer et al. [199] the focus was on utilizing computational methods to design and evaluate potential inhibitors for breast cancer, specifically D-glucofuranose and its derivatives. Through the application of quantum calculations, molecular docking, ADMET analysis, and SAR analysis, the researchers investigated the binding affinity of these compounds against two types of breast cancer proteases: cancer protease (3hb5) and triple-negative breast cancer protease (4pv5). Furthermore, molecular dynamics simulations were employed to validate the results obtained from docking experiments. Among the various compounds tested, three showed promising characteristics as potential drugs. These compounds, labeled as 03, 05, and 08, demonstrated high binding energy, non-toxicity, non-carcinogenicity, and excellent solubility in biological systems. In addition, the drug candidates satisfied the Lipinski rule and exhibited drug-likeness features. Considering their favorable properties, compounds 03, 05, and 08 hold the potential to be developed as standard drugs for breast cancer treatment. If successfully brought to the commercial market, these compounds could offer enhanced safety and reduced side effects compared to existing chemotherapy medications.

The article by Dushanan et al. [200] explores the efficacy of histone deacetylase (HDAC) inhibitors in cancer treatment using molecular dynamics simulations. The study focuses on vorinostat, panobinostat, abexinostat, belinostat, resminostat, dacinostat, and pracinostat. The researchers evaluated the stability of the HDAC enzyme and its interactions with these inhibitors. Results showed that vorinostat, panobinostat, and abexinostat exhibited higher stability. The study suggests that panobinostat and abexinostat could be potential lead compounds for HDAC inhibition in clinical practice and aid in the discovery of new inhibitors for further research. The study concludes that among the tested HDAC inhibitors, vorinostat, panobinostat, and abexinostat demonstrated higher stability in their complexes with the HDLP enzyme. These inhibitors showed low fluctuations in the enzyme's structure and strong binding interactions. The binding energies and interaction patterns also supported their efficacy. Panobinostat and abexinostat were identified as lead compounds for HDLP inhibition, alongside vorinostat. Pracinostat also showed potential as an inhibitor. The findings contribute to assessing the effectiveness of new HDAC inhibitors, reducing the time, costs, and resources required for clinical trials, and guiding the design of more potent inhibitors in the future.

# Conclusion

In conclusion, molecular dynamics simulations have emerged as a powerful tool for investigating the complex biological processes underlying cancer development and progression, as well as for the discovery and design of novel anti-cancer therapeutics. The use of molecular dynamics simulations in cancer research has provided valuable insights into the molecular mechanisms of oncogenesis, drug resistance, and the design of effective anti-cancer agents. Through the use of molecular dynamics simulations, researchers have been able to study the molecular interactions between anti-cancer agents and their targets at an atomistic level, which has facilitated the development of more potent and selective drugs. Moreover, the integration of molecular dynamics simulations with other computational methods, such as virtual screening and molecular docking, has enabled the efficient screening of large chemical libraries for the identification of novel anti-cancer drug candidates. The application of molecular dynamics simulations in combination with experimental techniques has led to the development of several promising anti-cancer agents, some of which are currently undergoing clinical trials. Despite the significant progress that has been made in the use of molecular dynamics simulations in cancer research, there are still several challenges that need to be addressed. These include the need for more accurate force fields and better sampling techniques to improve the accuracy and reliability of molecular dynamics simulations. In addition, there is a need for greater collaboration between computational and experimental researchers to validate the predictions of molecular dynamics simulations and to translate the findings into clinical practice. Overall, the use of molecular dynamics simulations in cancer research and drug discovery holds great promise for the development of more effective and targeted anti-cancer therapies. As computational power and modeling techniques continue to advance, molecular dynamics simulations are expected to play an increasingly important role in cancer research and drug development in the years to come.

88. Bernardi, R.C., M.C. Melo, and K. Schulten, Enhanced sampling techniques in molecular dynamics simulations of biological systems. Biochimica et Biophysica Acta (BBA)-General Subjects, 2015. 1850(5): p. 872-877.
89. Harpole, T.J. and L. Delemotte, Conformational landscapes of membrane proteins delineated by enhanced sampling molecular dynamics simulations. Biochimica Et Biophysica Acta (BBA)-Biomembranes, 2018. 1860(4): p. 909-926.
90. Schwantes, C.R., R.T. McGibbon, and V.S. Pande, Perspective: Markov models for long-timescale biomolecular dynamics. The Journal of chemical physics, 2014. 141(9).
91. Timpson, P., et al., Quantitative real-time imaging of molecular dynamics during cancer cell invasion and metastasis in vivo. Cell adhesion & migration, 2009. 3(4): p. 351-354.
92. Tripathi, S., G. Srivastava, and A. Sharma, Molecular dynamics simulation and free energy landscape methods in probing L215H, L217R and L225M βI-tubulin mutations causing paclitaxel resistance in cancer cells. Biochemical and biophysical research communications, 2016. 476(4): p. 273-279.
93. Dhanjal, J.K., et al., Embelin inhibits TNF-α converting enzyme and cancer cell metastasis: molecular dynamics and experimental evidence. BMC cancer, 2014. 14(1): p. 1-12.
94. Tang, Z.m., et al., Biodegradable nanoprodrugs:"delivering" ROS to cancer cells for molecular dynamic therapy. Advanced Materials, 2020. 32(4): p. 1904011.
95. Raffaini, G. and F. Ganazzoli, A Molecular Dynamics Study of a Photodynamic Sensitizer for Cancer Cells: Inclusion Complexes of γ-Cyclodextrins with C70. International Journal of Molecular Sciences, 2019. 20(19): p. 4831.
96. Martínez-Muñoz, A., et al., Selection of a GPER1 ligand via ligand-based virtual screening coupled to molecular dynamics simulations and its anti-proliferative effects on breast cancer cells. Anti-Cancer Agents in Medicinal Chemistry (Formerly Current Medicinal Chemistry-Anti-Cancer Agents), 2018. 18(11): p. 1629-1638.
97. Gade, D.R., et al., Elucidation of chemosensitization effect of acridones in cancer cell lines: Combined pharmacophore modeling, 3D QSAR, and molecular dynamics studies. Computational biology and chemistry, 2018. 74: p. 63-75.
98. Hirano, R., et al., Molecular mechanism underlying the selective attack of trehalose lipids on cancer cells as revealed by coarse-grained molecular dynamics simulations. Biochemistry and Biophysics Reports, 2021. 25: p. 100913.
99. Yang, S., et al., ReaxFF-based molecular dynamics simulation of DNA molecules destruction in cancer cells by plasma ROS. Physics of Plasmas, 2019. 26(8): p. 083504.
100. Zaccai, G., Molecular dynamics in cells: A neutron view. Biochimica et Biophysica Acta (BBA)-General Subjects, 2020. 1864(3): p. 129475.
101. Kashab, A.A.K., et al., Investigation of the effect of external force and initial pressure on the stability of cancer cells using molecular dynamics simulation. The European Physical Journal Plus, 2022. 137(8): p. 952.
102. Nguyen, H.L., et al., Elastic moduli of normal and cancer cell membranes revealed by molecular dynamics simulations. Physical Chemistry Chemical Physics, 2022. 24(10): p. 6225-6237.
103. Kordzadeh, A., M. Zarif, and S. Amjad-Iranagh, Molecular dynamics insight of interaction between the functionalized-carbon nanotube and cancerous cell membrane in doxorubicin delivery. Computer Methods and Programs in Biomedicine, 2023. 230: p. 107332.
104. Rivel, T., C. Ramseyer, and S. Yesylevskyy, Permeation of cisplatin through the membranes of normal and cancer cells: a molecular dynamics study. BioRxiv, 2018: p. 375980.
105. Tabrez, S., et al., Targeting glutaminase by natural compounds: Structure-based virtual screening and molecular dynamics simulation approach to suppress cancer progression. Molecules, 2022. 27(15): p. 5042.
106. Dang, T. and J. Chai, Molecular dynamics in esophageal adenocarcinoma: who's in control? Current Cancer Drug Targets, 2020. 20(10): p. 789-801.
45